\documentclass[aps,pra,amsmath,amsfonts,amssymb,twocolumn,nofootinbib]{revtex4}

\usepackage{amssymb}
\usepackage{graphicx,psfrag,color}
\usepackage{dcolumn}
\usepackage{bm}
\usepackage{mathbbol}
\usepackage[T1]{fontenc}

\newcommand{\med}[1]{\left\langle #1 \right\rangle}

\begin{document}
\flushbottom

\title{Finite temperature detection of quantum critical points: 
a comparative study}

\author{G. A. P. Ribeiro}
\author{Gustavo Rigolin}
\email{rigolin@ufscar.br}
\affiliation{Departamento de F\'isica, Universidade Federal de
S\~ao Carlos, 13565-905, S\~ao Carlos, SP, Brazil}

\date{\today}

\begin{abstract}
We comparatively study three of the most useful quantum information tools 
to detect quantum critical points (QCPs) when only finite temperature 
data are available. We investigate quantitatively how the 
quantum discord, the quantum teleportation based QCP detectors, 
and the quantum coherence spectrum pinpoint the QCPs 
of several spin-$1/2$ chains. We 
work in the thermodynamic limit (infinite number of spins) and with 
the spin chains in equilibrium with a thermal reservoir at temperature $T$. The models here studied are the $XXZ$ model with and without an 
external longitudinal magnetic field, the Ising transverse model, and the $XY$ model subjected to an external transverse magnetic field.
\end{abstract}


\maketitle

\section{Introduction}

The use of quantum information based tools to characterize a quantum
phase transition (QPT) brought to light the existence of genuine quantum correlations during a QPT \cite{ost02,osb02,wu04,ver04,oli06,dil08,sar08,fan14,gir14,hot08,hot09,ike23a,ike23b,ike23c,ike23d}. 
A QPT is characterized by a drastic 
change in the ground state describing a macroscopic system 
while we modify the system's Hamiltonian \cite{sac99,gre02,gan10,row10}. Traditionally, to properly observe a QPT,
we need to reduce the system's temperature such that the thermal fluctuations become small enough to not excite the system away from 
its ground state. In this scenario the system can be considered for 
all practical purposes at the absolute zero temperature ($T=0$) and 
we can be reassured that all measurements give information about the system's ground state alone. 

When the temperature $T$ is high enough, the probability to find the system in one of its excited states is no longer negligible. In this case the analysis of a QPT, a genuine feature of the system's ground state, 
is more subtle. Some tools may not
work at all, furnishing no clue to the existence of a quantum critical point (QCP) for the ground state. For instance,
the entanglement of formation \cite{woo98} between two spins is zero 
for some models in the vicinity of the QCP if the system is 
above a certain temperature \cite{wer10,wer10b}.\footnote{Note
that the same conclusion applies to the concurrence ($C$), 
an important entanglement monotone \cite{woo98}. This is true because the entanglement of formation ($EoF$) is a monotonically increasing function of $C$ and $EoF$ is zero if, and only if, $C=0$ \cite{woo98}.} 

Fortunately, there are quantum information tools that still allow 
us to infer the correct location of a QCP when only finite $T$
data are at hand. Our main goal here is to comparatively study 
the efficacy of the most promising tools to detect QCPs with finite
$T$ data. Specifically, we will study the three tools that 
stand out in this scenario, namely, the thermal quantum discord 
\cite{wer10b}, the quantum coherence spectrum \cite{li20}, and the teleportation based QCP detectors \cite{pav23,pav24}. The first 
tool is the quantum discord (QD) \cite{oll01,hen01} computed for systems
in thermal equilibrium \cite{wer10}, the second tool
is a spin-off of the quantum coherence (QC) \cite{wig63,fan14,gir14}, 
and the third one is based on the quantum teleportation 
protocol \cite{ben93,yeo02,rig17}.

It is worth mentioning that the comparison among the tools described 
above does not take into account their computational complexity. Furthermore, we do not take into account their operational meaning 
and experimental feasibility either. 

Indeed, the computation of the QD is not an easy task and for high
spins it is extremely difficult \cite{mal16}. The reason for that is 
related to how the computational resources needed to its evaluation 
scale as we increase the size of the system under investigation 
(QD is an NP-complete problem \cite{hua14}). Also, QD does not have
a direct experimental meaning and so far no general method to 
its direct measurement is available. A similar analysis
applies to the spectrum of the QC \cite{li20}. Its computation is also
resource intensive and no direct way for measuring it is available \cite{pav23}. The third set of tools, namely, the quantum teleportation
based QCP detectors, does not suffer from those problems, having a 
direct experimental meaning and being amenable to theoretical 
analysis for high spin systems \cite{pav23,pav24}.

\section{The critical point detectors}
\label{cpd}

The key ingredient needed to theoretically compute the following quantum 
information based QCP detectors is a two-qubit density matrix.
This density matrix completely characterizes a pair of nearest neighbor spins within the spin chain and it is obtained by tracing out all
but these two spins from the canonical ensemble density matrix describing the whole chain. In Sec. \ref{models} we will come back to this point 
providing further details. However, the important point now is that 
for all the models investigated in this work, this two-qubit density 
matrix has the following form \cite{wer10b,pav23,pav24,osb02},
\begin{equation}
\rho_{23}  = \left(
\begin{array}{cccc}
a & 0 & 0 & e\\
0 & b  & c & 0 \\
0 & c & b & 0 \\
e & 0 & 0 & d\\
\end{array}
\right), 
\label{rho23}
\end{equation}
where $a,b,c,d,e$ are real numbers.\footnote{Note that the
particular form of $\rho_{23}$ does not affect the 
application of the teleportation based QCP detectors. For this
set of QCP detectors, the calculations are straightforward and 
can be carried out analytically for an arbitrary two-qubit state
\cite{pav23,pav24}. For QD, though, if $\rho_{23}$ 
is not an ``X-state'', we cannot explicitly solve 
the associated optimization problem that gives QD anymore. In 
this case, we have to rely on numerical algorithms to obtain the quantum discord \cite{mal16,hua14}. Also, the present tools
can equally be applied to non-neighboring (distant) spins or more
than two or three spins. However, for computational constraints,
we restricted our analysis to the minimal number of spins needed
to apply each tool.}
The values of these numbers depend
on the particular model and on the temperature of the heat bath. In
Eq.~(\ref{rho23}), the subscripts $2$ and $3$ denote any two nearest 
neighbor spins while we reserve the number $1$ to represent an
extra qubit from or outside the spin chain that 
is teleported from Alice (qubit $2$) to Bob (qubit $3$) [see Sec. \ref{QCPdetectors}].  

\subsection{Thermal quantum discord}
\label{qd}

The QD aims to capture all genuine quantum correlations between two
physical systems (a bipartite system). It is defined as the difference
between two non-equivalent ways of extending to the quantum realm 
the classical mutual information between a bipartite system \cite{oll01,hen01}. For the density matrix (\ref{rho23}), the QD 
is \cite{cil10,hua13,yur15}
\begin{equation}
\mathcal{QD} =  S(\rho_3) - S(\rho_{23}) 
+ \min_{\theta\in [0,\frac{\pi}{2}]}\tilde{S}(\theta), \label{QD} 
\end{equation}
where $\rho_3=\mbox{Tr}_2{\rho_{23}}$ is the reduced state describing
qubit $3$, obtained after tracing out qubit $2$ from $\rho_{23}$, $S(\rho_3)$ and $S(\rho_{23})$ are, respectively, the von Neumann entropy 
of the states $\rho_3$ and $\rho_{23}$,  
%
\begin{eqnarray}
S(\rho_3) \!&\!=\!&\! -(a + b) \ln(a + b) - (b + d) \ln(b + d), \\ \label{s3} 
S(\rho_{23}) \!&\!=\!&\!-\frac{a+d-\sqrt{(a-d)^2+4 e^2}}{2} \nonumber \\
&&\times \ln\left(\frac{a+d-\sqrt{(a-d)^2+4 e^2}}{2}\right) \nonumber \\
&&-\frac{a+d+\sqrt{(a-d)^2+4 e^2}}{2} \nonumber \\
&&\times \ln\left(\frac{a+d+\sqrt{(a-d)^2+4 e^2}}{2}\right) \nonumber \\
&&-\!\left(b-c\right) \ln\!\left(b-c\right)
\!-\!\left(b+c\right) \ln\! \left(b+c\right)\!, \label{s23}
\end{eqnarray}
%
and 
\begin{equation}
\tilde{S}(\theta) = \Lambda_1\ln\Lambda_1 + \Lambda_2\ln\Lambda_2
-\sum_{j=1}^4\lambda_j\ln\lambda_j. \label{stilde}
\end{equation}
In Eq.~(\ref{stilde}) we have
\begin{eqnarray}
\Lambda_{1,2} \!&\! =\! &\! \frac{1}{2} [1\pm (a-d) \cos\theta], \\
\lambda_{1,2} \!&\! =\! &\! \frac{1}{4} (1+(a-d) \cos\theta \nonumber \\
\!&\! \pm\!\! &\!\! \left. 
\hspace{-.05cm}\sqrt{\![a-d+(\!a-2 b+d)\cos\theta]^2\!+\!4(|c|+|e|)^2 \sin^2\!\theta}\right)\!\!,
\nonumber \\
&  &  \\
\lambda_{3,4} \!&\! =\! &\! \frac{1}{4} (1-(a-d) \cos\theta \nonumber \\
\!&\! \pm\!\! &\!\! \left. 
\hspace{-.05cm}\sqrt{\![a-d-(\!a-2 b+d)\cos\theta]^2\!+\!4(|c|+|e|)^2 \sin^2\!\theta}\right)\!\!.
\nonumber \\
& & 
\end{eqnarray}

Note that in the previous expressions ``$\ln$'' is the natural logarithm
and the minimization of $\tilde{S}(\theta)$ must be implemented 
numerically. Once we have the two-qubit density matrix as given by 
Eq.~(\ref{rho23}), we compute Eq.~(\ref{stilde}) and numerically 
search for its minimum value assuming it is a function of 
$\theta$, with $\theta\in [0,\pi/2]$. 

\subsection{Quantum coherence spectrum}

The QC spectrum \cite{li20} is actually two different quantities
defined to investigate the spectrum of the operator defining the
QC \cite{fan14,gir14}. The QC studied here \cite{gir14} 
is a simplified version of the Wigner-Yanase skew information \cite{wig63}, which aims at quantifying the amount of information
a density matrix contains with respect to an observable, in particular
when the latter does not commute with the density matrix.
Note that by its very definition, the QC is observable dependent.
QC is also related to an interesting extension of the Heisenberg
uncertainty relation for mixed states \cite{luo05} or to the 
quantification of the coherence of a quantum state \cite{gir14}.

In its more experimentally friendly version and for a two-qubit
density matrix, QC is defined as 
\cite{gir14},
\begin{equation}
\mathcal{QC}(K) = -\frac{1}{4}\mbox{Tr}\{[\rho_{23},K]^2\}, \label{QC}
\end{equation}
where $\mbox{Tr}$ denotes the trace operation, $[A,B]=AB-BA$, and $K$
is a $4\times 4$ matrix representing any observable associated with
a two-qubit system. Note that we are highlighting in the definition 
above the dependence of QC on the observable $K$. 

If one computes the spectrum of $[\rho_{23},K]^2$, 
namely, determines its four eigenvalues $\alpha_1, \ldots, \alpha_4$, 
one can define the following two quantities \cite{li20}, 
\begin{eqnarray}
\mathcal{S}_{QC}^{K} &=& - \sum_{n=1}^4|\alpha_n|\ln|\alpha_n|, \label{sqc} \\
\mathcal{L}_{QC}^{K} &=& - \sum_{n=1}^4\ln|\alpha_n|. \label{lqc}
\end{eqnarray}
The superscripts in Eqs.~(\ref{sqc}) and (\ref{lqc}) remind us that 
they both depend on the observable $K$. The first quantity, 
$\mathcal{S}_{QC}^{K}$, is called coherence entropy and the second one,
$\mathcal{L}_{QC}^{K}$, logarithm of the spectrum \cite{li20}.  
We should also note that in Ref.
\cite{li20} the above quantities were defined without taking the 
moduli of the eigenvalues. This is inconsistent since, 
as we will show next, these eigenvalues can become negative for certain observables if we use the density matrix (\ref{rho23}). In other words,
we must take the absolute value of $\alpha_n$, as in Eqs. (\ref{sqc}) and (\ref{lqc}), if we want the logarithm to be a well defined real function.
Furthermore, the second quantity above,
$\mathcal{L}_{QC}^{K}$, even when defined with $|\alpha_n|$ instead of 
$\alpha_n$, continues to be ill-defined and problematic. The reason for 
that is related to the fact that there are certain combinations of 
$\rho_{23}$ and $K$ that lead to at least one $\alpha_n$
being zero. 
And if $\alpha_n$ is zero $\ln|\alpha_n|$ is not defined 
($\lim_{x\rightarrow 0}\ln |x|\rightarrow -\infty$). 
On the other hand,
$\mathcal{S}_{QC}^{K}$ is perfectly legitimate since $\lim_{x\rightarrow 0}|x|\ln |x| = 0$. For a clear and simple illustration of this point, see the discussion around Eq.~(\ref{bell19}).

We will provide in Sec. \ref{models} a couple of examples where at least
one $\alpha_n$ is zero. It will 
turn out that the ``robustness'' of $\mathcal{L}_{QC}^{K}$ to detect
QCPs using extremely high $T$ data, as reported in Ref. \cite{li20},
is a consequence of its faulty definition. This alleged
high temperature robustness is not even restricted to the QCPs either. 
In many cases $\alpha_n$ is zero in regions of the parameter space 
defining the Hamiltonian where no QPT is 
taking place. Also, whenever 
$\alpha_n$ becomes zero, at or away from a QCP, this feature is independent of the 
value of the system's temperature and is a consequence of a particular
symmetry of the system's Hamiltonian. This is the case for all the models 
investigated in ref. \cite{li20} and where $\mathcal{L}_{QC}^{K}$ was 
shown to be insensitive to temperature increases 
(see Sec. \ref{models}).

Following Ref. \cite{li20}, we restrict our analysis to
three local observables $K$, namely, 
$K=\mathbb{1}\otimes \sigma^x,\mathbb{1}\otimes \sigma^y$, and
$\mathbb{1}\otimes \sigma^z$, where $\mathbb{1}$ is the $2\times 2$
identity matrix acting on qubit 2 (Alice) and 
$\sigma^x,\sigma^y$, and $\sigma^z$ are the standard Pauli matrices
acting on qubit 3 (Bob). 
The matrix representation of the two-qubit state $\rho_{23}$
[see Eq.~(\ref{rho23})] 
is given in the computational basis 
$\{|00\rangle,|01\rangle,|10\rangle,|11\rangle\}$, where $\sigma^z$
is diagonal.

A direct calculation using Eq.~(\ref{rho23}) and the representation 
of $\mathbb{1}\otimes\sigma^x$ in the basis where $\sigma^z$
is diagonal leads to the following eigenvalues for
$[\rho_{23},\mathbb{1}\otimes\sigma^x]^2$,
\begin{eqnarray}
\alpha_{1,2}^x &=&\alpha(1,1),  \label{a1x}\\
\alpha_{3,4}^x &=&\alpha(-1,1), \label{a3x}
\end{eqnarray}
where
\begin{eqnarray}
\alpha(\epsilon_1,\epsilon_2) &=& -\frac{1}{2} \left[ (a-b)^2+(b-d)^2+2 
(c-\epsilon_2 e)^2
\right. \nonumber \\
&&+\left.\epsilon_1  (a-2 b+d) \sqrt{(a-d)^2+4(c-\epsilon_2 e)^2}\right].
\nonumber \\
&& \, \label{a}
\end{eqnarray}

Similarly, for $[\rho_{23},\mathbb{1}\otimes\sigma^y]^2$ we have
\begin{eqnarray}
\alpha_{1,2}^y &=&\alpha(1,-1),  \label{a1y}\\
\alpha_{3,4}^y &=&\alpha(-1,-1), \label{a3y}
\end{eqnarray}
and for $[\rho_{23},\mathbb{1}\otimes\sigma^z]^2$ we get
\begin{eqnarray}
\alpha_{1,2}^z &=&-4c^2, \label{a1z}\\
\alpha_{3,4}^z &=&-4e^2. \label{a3z}
\end{eqnarray}
Note that the superscripts in the eigenvalues above mark the 
corresponding operator $K$ that we used in each one of the three previous
calculations. Also, all eigenvalues are doubly degenerate.

If we look at the eigenvalues given by Eqs.~(\ref{a1z}) and (\ref{a3z}),
we clearly see that they are all negative. This proves that we must 
define Eqs.~(\ref{sqc}) and (\ref{lqc}) using the magnitude of those eigenvalues. We should not use the eigenvalues directly, 
as was done in Ref. \cite{li20}. In Eqs.~(\ref{a1x}) and 
(\ref{a3x}) or in Eqs.~(\ref{a1y}) and (\ref{a3y}) we also have 
that at least two out of four eigenvalues are clearly negative.  
This is true because a pair of degenerate eigenvalues is 
given by either $-u + v$ or $-u - v$, 
with $u$ and $v$ positive numbers. And whenever 
$v<u$ all four eigenvalues become negative.

Before we move on, we show a very simple case where
we have two out of four eigenvalues zero. This happens for  
all the operators $K$ that we use here, proving that 
$\mathcal{L}_{QC}^{K}$, Eq.~(\ref{lqc}), cannot be defined for 
all two-qubit states.  

Let us take the following Bell state, namely, 
$|\Phi^+\rangle=(|00\rangle + |11\rangle)/\sqrt{2}$. Its density matrix
is
\begin{equation}
\rho_{\Phi^+} = |\Phi^+\rangle \langle\Phi^+| = 
\left(
\begin{array}{cccc}
\frac{1}{2} & 0 & 0 & \frac{1}{2} \\
0 & 0 & 0 & 0 \\
0 & 0 & 0 & 0 \\
\frac{1}{2} & 0 & 0 & \frac{1}{2}
\end{array}
\right).\label{bell19}
\end{equation}
Comparing with Eq.~(\ref{rho23}), we get $a=d=e=1/2$ and 
$b=c=0$. Therefore, Eqs.~(\ref{a1x})-(\ref{a3z}) become
\begin{eqnarray}
\alpha_{1,2}^x = \alpha_{1,2}^y = \alpha_{3,4}^z &=&-1, \\
\alpha_{3,4}^x = \alpha_{3,4}^y = \alpha_{1,2}^z &=&0.
\end{eqnarray}
The above result is not restricted to this particular Bell state.
The same is true for the other three. Moreover, the existence of 
null eigenvalues is not an exclusive feature
of an entangled state, such as the Bell state above.  
If we employ, for instance, the separable states $|00\rangle$ or 
$|11\rangle$, we also get a pair of null eigenvalues.

\subsection{Teleportation based QCP detectors}
\label{QCPdetectors}

The teleportation based QCP detectors \cite{pav23,pav24}
use a pair of qubits ($\rho_{23}$) 
from a spin chain as the quantum resource 
(quantum communication channel) through which the standard 
teleportation protocol \cite{ben93} is implemented. 
Since a QPT induces a drastic change in the system's ground state, 
it is expected that the state describing this 
pair of qubits also changes substantially. 
This change will eventually affect the efficiency of the teleportation protocol. As such, an abrupt change in the efficiency of the 
teleportation protocol may indicate a QPT and the exact location 
of the corresponding QCP \cite{pav23,pav24}. 

In general, the state describing a pair of qubits from a spin chain is
a mixed state. Thus, to properly construct the teleportation based QCP
detectors, we need to recast the standard teleportation protocol in
the formalism of density matrices \cite{pav23,pav24,rig17,rig15}. 

\begin{figure}[!ht]
\includegraphics[width=8cm]{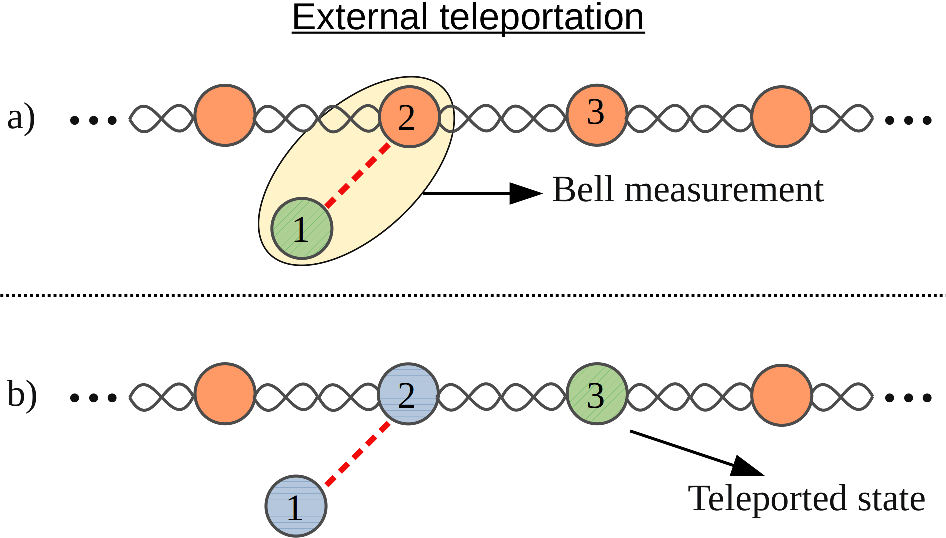}
\caption{\label{fig_scheme}(color online) 
A single run of the external 
teleportation protocol using a pair of qubits 
from a spin chain is described by the following steps.
Alice and Bob agree on which pair of qubits to use as the 
quantum resource to implement the teleportation protocol. This pair 
is illustrated  by qubits $2$ and $3$ in the figure. Alice brings 
an external qubit to be teleported to Bob.
This is qubit $1$ depicted in the figure. Subsequently Alice projects
qubits $1$ and $2$ onto a Bell state. 
This is a standard Bell measurement (BM).
See panel (a) above. Alice then tells Bob of her BM result by
sending two classical bits to Bob since there are four possible outcomes
after a BM. Finally, based on the information received by Bob from
Alice, he implements a corresponding unitary operation on his qubit 
to finish the protocol. 
This step is represented in the panel (b) above.}
\end{figure}

\begin{figure}[!ht]
\includegraphics[width=8cm]{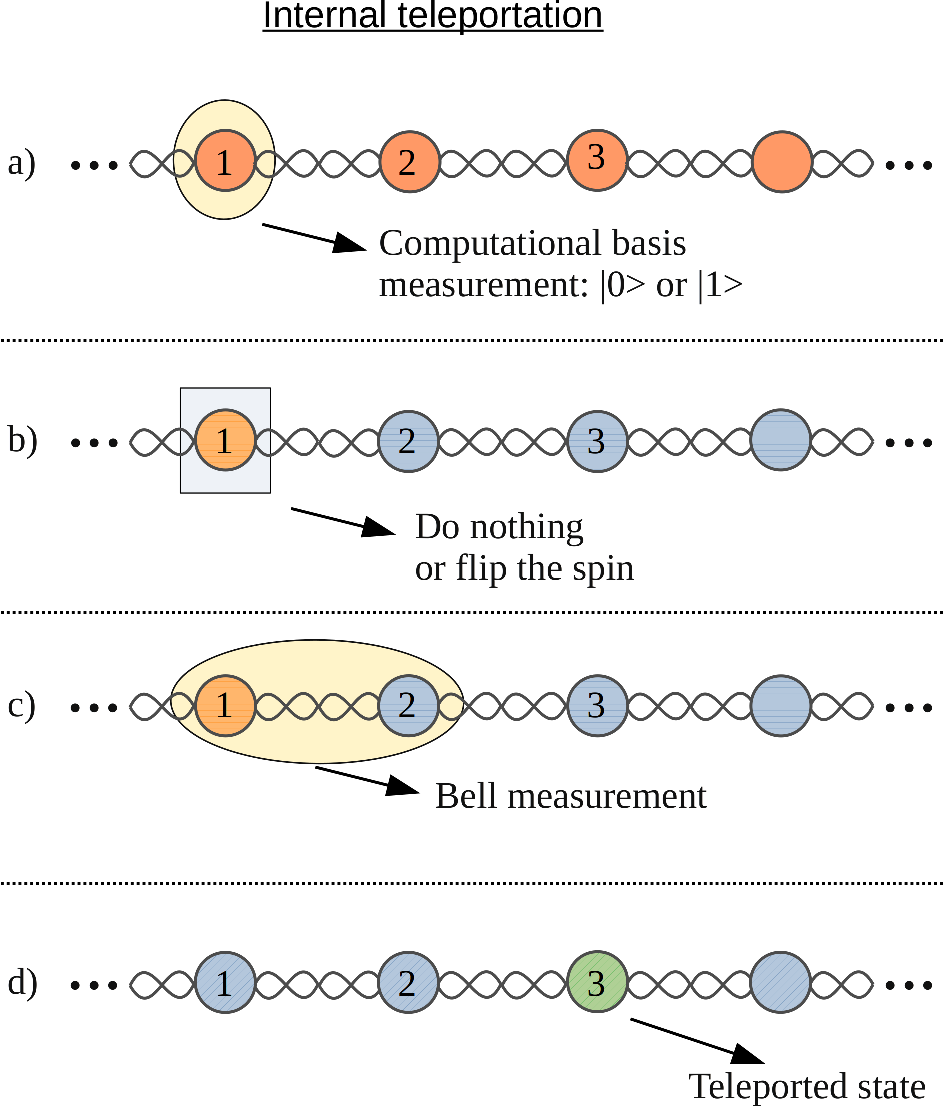}
\caption{\label{fig_scheme2}(color online) 
A single run of the internal teleportation protocol 
using a pair of qubits from a spin chain is described by the following steps. Alice projects the qubit $1$ onto the computational basis [panel (a)]. 
Then, she either applies 
onto it the spin flip operation ($\sigma_1^x$) or 
does nothing [panel (b)] according to recipe given in 
Ref. \cite{pav24}. Finally, Alice and 
Bob execute the standard teleportation protocol as explained in 
Fig. \ref{fig_scheme} [panels (c) and (d)].}
\end{figure}

Qubits $2$ and $3$, described by $\rho_{23}$, 
constitute the quantum resource shared by Alice (qubit $2$) and Bob 
(qubit $3$). It is obtained tracing out from the whole chain all 
but these two qubits. The qubit to be teleported or input qubit 
can be an external qubit from the chain \cite{pav23} or another qubit 
from the spin chain \cite{pav24}. 
In both cases it is formally described by the density matrix $\rho_1$. 
If the input qubit does not belong to the chain, we have the 
external teleportation based QCP detector while if the input belongs
to the chain we have the internal teleportation based QCP detector.
In Figs.~\ref{fig_scheme} and \ref{fig_scheme2}
we schematically show how the two approaches work.

For the external teleportation protocol, the density matrix describing the three qubits before the 
teleportation begins is \cite{pav23}
\begin{equation}
 \rho = \rho_{1} \otimes \rho_{23},
\label{stepA}
\end{equation}
where $\rho_1$ is an arbitrary pure state that Alice can 
freely choose and $\rho_{23}$ is the density matrix describing
a pair of qubits from the spin chain.
For the internal teleportation protocol, the state describing the three qubits is in general given by $\rho_{123}$ \cite{pav24}, 
where the latter is obtained by tracing out all but qubits $1,2$,
and $3$ from the state describing the whole chain. 
In order to effectively obtain 
Eq.~(\ref{stepA}) in the internal teleportation protocol, 
Alice has to implement steps (a) and (b)
described in Fig. \ref{fig_scheme2} before starting the 
teleportation protocol. See ref. \cite{pav24} for all the details
of how this can be accomplished. At the end, Alice and
Bob will share an ensemble of states effectively given 
by Eq.~(\ref{stepA}), where $\rho_{23}$ is the density matrix describing a pair of qubits from the spin chain and 
$\rho_1$ is the density matrix associated with a single spin
from the chain.

At the end of one run of the teleportation protocol, as described in 
Figs.~\ref{fig_scheme} and \ref{fig_scheme2}, qubit $3$ with Bob is \cite{pav23,pav24,rig17}
\begin{equation}
\rho_{_{B_j}}=   \frac{U_jTr_{12}[P_j \rho P_j]U_j^\dagger}{Q_j}.
\label{stepD}
\end{equation}
In Eq.~(\ref{stepD}), $U_j$ is the unitary operation that Bob 
applies on his spin after being informed from Alice which Bell state 
$j$ she measured. The Bell measurement 
(BM) implemented by Alice projects qubits $1$ and $2$ onto one 
of the four Bell states.  
Also, $Tr_{12}$ denotes the partial trace over Alice's 
spins (qubits 1 and 2) and $P_j$ is one of the four projectors 
related to a BM,
\begin{eqnarray}
P_{\Psi^{\pm}} &=& |\Psi^{\pm}\rangle \langle \Psi^{\pm}|, \label{projectorA}\\  
P_{\Phi^{\pm}} &=& |\Phi^{\pm}\rangle \langle \Psi^{\pm}|. \label{projectorB}  
\end{eqnarray}
The Bell states are given by
\begin{eqnarray}
|\Psi^{\pm}\rangle&=&(|01\rangle \pm |10\rangle)/\sqrt{2},\label{BellA} \\ 
|\Phi^{\pm}\rangle&=&(|00\rangle \pm |11\rangle)/\sqrt{2}. \label{BellB}
\end{eqnarray}
The denominator in Eq.~(\ref{stepD}) gives the probability of Alice 
measuring the Bell state $j$ \cite{pav23,rig17},
\begin{equation}
Q_j = Tr[{P_j \rho}].
\label{prob}
\end{equation}

We should note that the unitary operation $U_j$ that Bob applies 
on his qubit also depends on the type of quantum resource shared 
with Alice \cite{pav23,pav24,rig17}. In the standard teleportation protocol, 
where the state $\rho_{23}$ is a Bell state 
$|k\rangle$, with $k=\Psi^{\pm},\Phi^{\pm}$, we have that 
$U_j$ is given by the following set $S_k$ 
of four unitary operators,
\begin{eqnarray}
S_{\Phi^+}=\{U_{\Phi^+},U_{\Phi^-},U_{\Psi^+},U_{\Psi^-}\}
=\{\mathbb{1},\sigma^z,\sigma^x,\sigma^z\sigma^x\},
\label{ss1} \\
S_{\Phi^-}=\{U_{\Phi^+},U_{\Phi^-},U_{\Psi^+},U_{\Psi^-}\}
=\{\sigma^z,\mathbb{1},\sigma^z\sigma^x,\sigma^x\}, \label{ss2}\\
S_{\Psi^+}=\{U_{\Phi^+},U_{\Phi^-},U_{\Psi^+},U_{\Psi^-}\}=\{\sigma^x,\sigma^z\sigma^x,\mathbb{1},\sigma^z\}, \label{ss3}\\  
S_{\Psi^-}=\{U_{\Phi^+},U_{\Phi^-},U_{\Psi^+},U_{\Psi^-}\}=\{\sigma^z\sigma^x,\sigma^x,\sigma^z,\mathbb{1}\}.\label{ss4}
\end{eqnarray}

Here, the state $\rho_{23}$ is a mixed state that changes 
after a QPT. In one phase it is closer to one Bell state and
in another phase it is more similar to another one. 
Therefore, the teleportation based QCP detectors are defined  
by picking the optimal case out of the four sets $S_k$ above.

For the external teleportation based QCP detector, the fidelity \cite{uhl76,nie00} is employed to assess the efficiency of the teleportation protocol. 
The fidelity quantifies how close or similar 
two states are to each other. 
It is employed here to compare the similarity of 
the output state with Bob at the end of the protocol with the input 
state teleported by Alice. Since Alice always choose pure states to 
teleport to Bob, the fidelity becomes
\begin{equation}
F_j(|\psi \rangle,S_k) = \langle \psi | \rho_{_{B_j}} | \psi \rangle,
\label{Fidj}
\end{equation}
where $| \psi \rangle$ is any single qubit pure state external to the 
chain (see Fig.~\ref{fig_scheme}) while $\rho_{_{B_j}}$ is given by
Eq.~(\ref{stepD}). The fidelity is one if the two states are identical
and zero if they are orthogonal. 

After several runs of the protocol, each Bell state will be measured by 
Alice with probability $Q_j$. Thus, the relevant quantity in this case 
is the 
average fidelity \cite{pav23,gor06},
\begin{equation}
\overline{F}(|\psi\rangle,S_k)= \sum_{j=\Psi^{\mp},\Phi^{\mp}} 
Q_jF_j(| \psi \rangle,S_k). \label{Fbar}
\end{equation}
Optimizing over $|\psi\rangle$ (picking the maximum over all 
pure input states on the Bloch sphere) and over $S_k$ 
(the four sets of unitary corrections available to Bob), 
we get the maximum mean fidelity \cite{pav23},
\begin{equation}
\overline{\mathcal{F}} = \max_{\{|\psi\rangle,S_k\}}{\overline{F}(|\psi\rangle,S_k)}.
\label{fmax}
\end{equation}
Equation (\ref{fmax}) is the most accurate QCP detector based on 
the external teleportation protocol. 
For the density matrix $\rho_{23}$ given by
Eq.~(\ref{rho23}) we obtain
\begin{equation}
\overline{\mathcal{F}}_{ext} = \max\left[ 
2 b, 1-2 b,  \frac{1}{2} + |c| + |e|
\right],
\label{fmax2}
\end{equation}
where we used the normalization condition 
$\mbox{Tr}(\rho_{23})=a + 2 b + d=1$ to arrive at the expression above.
Note that we append the subscript ``ext'' to 
$\overline{\mathcal{F}}$ to make it clear that this
particular expression only applies to 
the external teleportation case.

We should note that $\overline{\mathcal{F}}_{ext}$ depends only on the 
two-qubit state $\rho_{23}$, similarly to
$\mathcal{QD}$, $\mathcal{S}^K_{QC}$, and $\mathcal{L}^K_{QC}$.
This means that once $\rho_{23}$ is measured or calculated, all these 
QCP detectors can be computed.
From an experimental point of view, however, 
$\overline{\mathcal{F}}_{ext}$
has a clear direct operational meaning. If we teleport a representative 
sample of pure qubits spanning the Bloch sphere, with Bob 
choosing randomly the set $S_k$ from which he picks the unitary correction to apply on his qubit,
and then compute the
corresponding mean fidelities, we have that 
$\overline{\mathcal{F}}_{ext}$ is given by the greatest
mean fidelity of all cases. The fidelity can be determined with the 
knowledge of Alice's input state
and Bob's output state at the end of the teleportation protocol. 
Alice's state is chosen by her and is readily known after she prepares
it. Bob's state can be experimentally determined after the 
teleportation is finished. Furthermore, the experimental
determination of Bob's state, a single qubit state, is accomplished 
by measuring one-point correlation functions (magnetization) alone 
\cite{pav23,pav24,nie00}. On the other hand, $\rho_{23}$ is determined
by measuring two-point correlation functions \cite{pav23,pav24,nie00}.

For the internal teleportation based QCP detector, 
the input state and the 
output state are mixed states and the fidelity becomes a really 
complicated expression \cite{pav24}. Therefore, we use the 
trace distance \cite{nie00,tos23,tos24} to quantify the similarity of the 
input and output states. In Ref. \cite{pav24} we also 
showed that in this case not only the trace distance is simpler but more sensitive 
to detect QCPs for all the models studied here. 

Alice's input state now is fixed and given by a single spin of the chain.
Since we are dealing with translational invariant spin chains,  
\begin{equation}
\rho_1=\rho_{2}=\mbox{Tr}_3(\rho_{23}) =  
\left(
\begin{array}{cc}
a+b & 0\\
0 & b+d 
\end{array}
\right).
\label{rho1}
\end{equation}

The trace distance between Bob's final state and Alice's input after a 
single run of the teleportation protocol is \cite{pav24}
\begin{equation}
D_j(S_k)=D(\rho_1,\rho_{_{B_j}}) = \frac{1}{2}\mbox{Tr}
\left|\rho_1 - \rho_{_{B_j}}\right|,
\label{Dj}
\end{equation}
where $\rho_{_{B_j}}$ and $\rho_1$ are given by Eqs.~(\ref{stepD}) and
(\ref{rho1}) and $|A| = \sqrt{A^\dagger A}$. 

The trace distance is half the Euclidean distance between 
the points on the Bloch sphere representing the two states above.
This means that two identical states have $D_j=0$  and 
orthogonal pure states have $D_j=1$, the maximum value for $D_j$.
For two single qubits we have \cite{nie00}
\begin{equation}
D_j(S_k) = \frac{1}{2}\sqrt{(\Delta r_x)^2+(\Delta r_y)^2+(\Delta r_z)^2},
\label{Dj2}
\end{equation}
where $\Delta r_\alpha(t) = \mbox{Tr}(\rho_1\sigma^\alpha)
-\mbox{Tr}(\rho_{_{B_j}}\sigma^\alpha)$. 

Similarly to the fidelity,
the mean trace distance after several runs of the teleportation 
protocol is
\begin{equation}
\overline{D}(S_k)= \sum_{j=\Psi^{\mp},\Phi^{\mp}} 
Q_jD_j(S_k). \label{Dbar}
\end{equation}
Since the more similar two states, the lower their trace distance, 
we now want the minimum over all sets $S_k$. As such, the 
internal teleportation QCP detector is \cite{pav24}
\begin{equation}
\overline{\mathcal{D}}_{int} = \min_{\{S_k\}}{\overline{D}(S_k)}=
|1 - 2 (b + d)|\min\left[1 - D_{-}, D_{+} \right],
\label{dmin}
\end{equation}
where
\begin{equation}
D_\pm = 2 b + d - (b + d)^2 \pm |(b + d)^2 - d|.
\end{equation}
To arrive at Eq.~(\ref{dmin}) we employed several properties of 
the density matrix $\rho_{23}$. We used that $0\leq a,b,d\leq 1$,
that $c$ and $e$ are real numbers, and the normalization condition $a+2b+d=1$. 

The experimental procedure to directly determine the minimum mean trace distance is akin to the one already given for the maximum mean fidelity. 
Now, however, we do not even need to cover the whole Bloch sphere since
the input state is always one spin of the chain, described by the same
state $\rho_1$ at every run of the teleportation protocol. The
experimental procedure to determine $\rho_1$ before the teleportation
begins and Bob's state at the end of one run of the protocol is
based on the experimental determination of one-point correlation functions, as already explained when we discussed the operational 
interpretation of the maximum 
mean fidelity \cite{pav23,pav24}.
With $\rho_1$, measured only once, and with Bob's final state, 
measured as already described after each run of the
internal teleportation protocol, all
relevant quantities to the determination of the minimum mean 
trace distance can be computed.

\section{The models}
\label{models}

The models investigated here are the local ones given in Ref. \cite{li20},
in particular those for which $\mathcal{L}^K_{QC}$ apparently beat 
the quantum discord in providing the exact location of the QCPs at 
finite $T$. We will be dealing with one dimensional translational 
invariant spin-1/2 chains 
in the thermodynamic limit, i.e., with $L\rightarrow \infty$, where 
$L$ represents the number of spins in the chain. They all satisfy 
periodic boundary conditions, namely, 
$\sigma^{x,y,z}_{L+1} = \sigma^{x,y,z}_1$. The subscripts 
in the Pauli matrices indicate on which qubit they act and
the spin chains are initially in equilibrium with a thermal reservoir at temperature $T$ (heat bath). 

The density matrix describing the chain of $L$ spins is the 
canonical ensemble density matrix, 
\begin{equation}
\varrho=\frac{e^{-H/kT}}{Z}, \label{canonical} 
\end{equation}
where $Z=Tr[e^{-H/kT}]$ is the partition function and the Boltzmann's constant is given by $k$.

If we trace out all but two nearest neighbor spins from the chain,
the density matrix describing them is given by Eq.~(\ref{rho23}). In terms of the one- and two-point correlation functions, we obtain \cite{wer10b,pav23,osb02} 
\begin{eqnarray}
a &=& \frac{1+2\med{\sigma^z_2}+\med{\sigma_2^z\sigma_{3}^z}}{4}, 
\label{varA}\\
b &=& \frac{1-\med{\sigma_2^z\sigma_{3}^z}}{4}, \\
c &=& \frac{\med{\sigma_2^x\sigma_{3}^x}+\med{\sigma_2^y\sigma_{3}^y}}{4}, \\
d &=&  \frac{1-2\med{\sigma^z_2}+\med{\sigma_2^z\sigma_{3}^z}}{4}, \\
e &=& \frac{\med{\sigma_2^x\sigma_{3}^x}-\med{\sigma_2^y\sigma_{3}^y}}{4},
\label{varE}
\end{eqnarray}
where for $s=x,y,z$ we have
\begin{eqnarray}
z&=&\med{\sigma_j^z}=Tr[\sigma_j^z\, \varrho], \label{z} \\
ss&=&\med{\sigma_j^s\sigma_{j+1}^s}=
Tr[\sigma_j^s\sigma_{j+1}^s\, \varrho]. \label{zz}
\end{eqnarray}
The details of those calculations in the thermodynamic limit 
($L \rightarrow \infty$) are in Refs. \cite{yan66,clo66,klu92,bor05,boo08,tri10,tak99,lie61,bar70,bar71,pfe70,zho10} and in Ref. \cite{wer10b} we review them in the present notation. 
In Ref. \cite{pav23} we also investigate how $\med{\sigma_j^z}$ and 
$\med{\sigma_j^s\sigma_{j+1}^s}$ behave for several values
of $T$ as we drive the system's Hamiltonian through its parameter space.

For the external teleportation based QCP detector, 
the density matrix describing 
Alice's input is $\rho_1^{ext}= |\psi\rangle \langle\psi |$, where
$|\psi\rangle=\cos(\theta/2)|0\rangle+\sin(\theta/2)e^{i\chi}|0\rangle$,
with $\theta \in [0,\pi]$ and $\chi\in [0,2\pi)$. Maximizing 
the mean fidelity (\ref{Fbar}) over $S_k$ and over all states on 
the Bloch sphere, i.e., maximizing over $\theta$ and $\chi$, gives the maximal mean fidelity $\overline{\mathcal{F}}_{ext}$ [Eq.~(\ref{fmax2})]. On the other hand, Alice's input for the 
internal teleportation based QCP detector is fixed by Eq.~(\ref{rho1}).
In this case, we only minimize the mean trace distance (\ref{Dbar}) 
over $S_k$ to obtain
the minimal mean trace distance $\overline{\mathcal{D}}_{int}$ 
[Eq.~(\ref{dmin})].

If we use Eqs.~(\ref{varA})-(\ref{varE}), Eq.~(\ref{fmax2}) becomes
\begin{equation}
\overline{\mathcal{F}}_{ext}
=\max{\left[\frac{1 + |xx|}{2}, \frac{1 + |yy|}{2}, 
\frac{1 + |zz|}{2}\right]}.
\label{fmax2a}
\end{equation}
To obtain Eq.~(\ref{fmax2a}), we used the following mathematical identity,
$\max[ |c| + |e|] = \max[|c + e|, |c - e|]$.

Similarly, using Eqs.~(\ref{varA})-(\ref{varE}) we can write  Eq.~(\ref{dmin}) as follows,
\begin{equation}
\overline{\mathcal{D}}_{int} =
\frac{1}{4}\left[(2 - |z^2 + zz|) |z| + |z^3 - z\cdot zz|\right],
\label{dmin2}
\end{equation}
where the dot between $z$ and $zz$ means the standard multiplication
between two real numbers. 

We should note that Eq.~(\ref{dmin2}) is not useful to analyze QPTs 
for spin chains with zero magnetization \cite{pav24}. 
Looking at Eq.~(\ref{dmin2}), we easily see that whenever $z=0$ 
we always have $\overline{\mathcal{D}}_{int} = 0$. 
This is why  
$\overline{\mathcal{D}}_{int}$ does not show up when we study the 
first case below.

\subsection{The XXZ model with no field}
\label{xxz0}

The Hamiltonian ($\hbar=1$) describing the XXZ model with no external
magnetic field is
\begin{equation}
H = \sum_{j=1}^{L}\left( \sigma^{x}_{j}\sigma^{x}_{j+1} +
\sigma^{y}_{j}\sigma^{y}_{j+1} + \Delta
\sigma^{z}_{j}\sigma^{z}_{j+1} \right). \label{Hxxz}
\end{equation}
The tuning parameter for this model is the anisotropy   
$\Delta$. At $T=0$ the XXZ model possesses two
QCPs \cite{tak99}. When $\Delta = -1$ a first-order QPT occurs and
the ground state changes from a ferromagnetic ($\Delta < -1$) to a critical antiferromagnetic phase ($-1 < \Delta < 1$). 
When $\Delta = 1$ a continuous QPT happens and the system enters 
an Ising-like antiferromagnet phase for $\Delta > 1$.

For the Hamiltonian (\ref{Hxxz}) we have that
$z=0$ and $xx=yy$. Therefore, the four eigenvalues used to
define $\mathcal{S}_{QC}^{x}$ and $\mathcal{L}_{QC}^{x}$ become
$\alpha_{1,2}^x=-(xx^2 + zz^2 + 2 zz |xx|)/4$ and 
$\alpha_{3,4}^x=-(xx^2 + zz^2 - 2 zz |xx|)/4$
[see Eqs.~(\ref{a1x}) and (\ref{a3x})]. If $xx\geq0$, we have 
$\alpha_{1,2}^x=-(xx + zz)^2/4$ and $\alpha_{3,4}^x=-(xx - zz)^2/4$.
If $xx\leq 0$, on the other hand, we have $\alpha_{1,2}^x=-(xx - zz)^2/4$ and $\alpha_{3,4}^x=-(xx + zz)^2/4$. Hence, 
without loss of generality, we fix our attention to the following
set of eigenvalues,
\begin{eqnarray}
\alpha_{1,2}^x&=&-(xx + zz)^2/4, \label{a12XXZ0} \\
\alpha_{3,4}^x&=&-(xx - zz)^2/4. \label{a34XXZ0}
\end{eqnarray}
Note that for this particular model 
$\mathcal{S}_{QC}^{x}=\mathcal{S}_{QC}^{y}$ and 
$\mathcal{L}_{QC}^{x} =\mathcal{L}_{QC}^{y}$ since the 
eigenvalues defining those quantities are the same. 
Also, for this model
the authors of Ref. \cite{li20} did not work with 
$\mathcal{S}_{QC}^{z}$ and $\mathcal{L}_{QC}^{z}$ and thus we will 
not work with them either.

The first thing worth mentioning is that all eigenvalues, Eqs.~(\ref{a12XXZ0})
and (\ref{a34XXZ0}), are negative. This is another case justifying that
one should always take the absolute values of those eigenvalues when
defining $\mathcal{S}_{QC}^K$ and $\mathcal{L}_{QC}^{K}$. Otherwise 
we would face logarithms with negative arguments. 

Second, if $|xx|=|zz|\neq 0$ we will always have two null eigenvalues. 
As such, $\mathcal{L}_{QC}^{x,y}$ will always be undefined (diverge)
in this scenario [see Eq.~(\ref{lqc})]. 
For the XXZ model with no field, the two QPTs occur exactly when this happens. When $\Delta=-1$ we have $xx=-zz$ and when
$\Delta=1$ we have $xx=zz$. Moreover, at $\Delta=\pm 1$ we will always
have $|xx|=|zz|$, no matter how high the temperature is. This 
is a consequence of specific symmetries of the Hamiltonian at 
those points.
For instance, when $\Delta= 1$ we have $H = \sum_{j=1}^{L}\left( \sigma^{x}_{j}\sigma^{x}_{j+1} + \sigma^{y}_{j}\sigma^{y}_{j+1} + \sigma^{z}_{j}\sigma^{z}_{j+1} \right)$ and it is obvious that
the two two-point correlation functions $xx$ and $zz$ should be equal
due to the rotational invariance of the Hamiltonian. Furthermore, 
this symmetry must be respected not only by the ground state but by 
any other excited state. Therefore, the canonical ensemble
density matrix 
describing the system at equilibrium with a heat bath also respects
it and we must always have $xx=zz$ for any $T$. A similar argument
also shows that $xx=-zz$ when $\Delta = -1$ for any $T$.

The above analysis explains why $\mathcal{L}_{QC}^{x}$ was
incorrectly considered robust against temperature increases in 
detecting the QCPs for
the XXZ model with no field \cite{li20}. This is a consequence of the divergence of $\mathcal{L}_{QC}^{x,y}$ at $\Delta=\pm1$ for any $T$ and
not of its unique ability to detect QCPs. As already stressed before,
$\mathcal{L}_{QC}^{K}$ should not be used when any of the eigenvalues 
appearing in its definition is zero. 

\begin{figure}[!ht]
\includegraphics[width=8cm]
{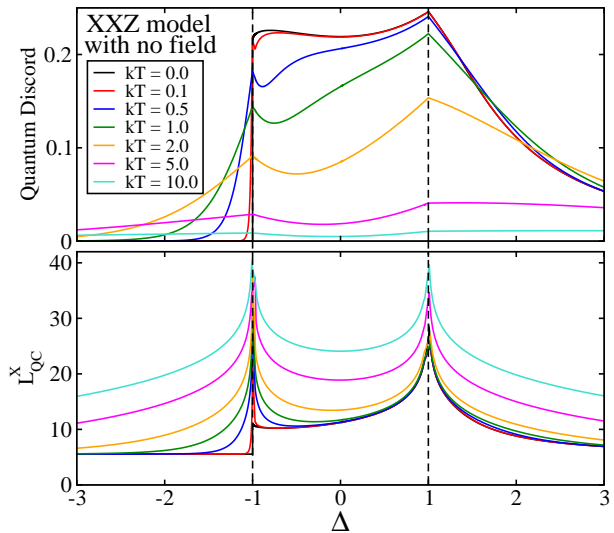}
\caption{\label{figQDLqcXXZ0}(color online)  
Quantum discord, Eq.~(\ref{QD}), 
and the logarithm of the spectrum, 
Eq.~(\ref{lqc}), as a function of $\Delta$ for several values of 
temperature. For the upper panel, the temperature increases from top
to bottom between the QCPs, 
while for the lower panel, it increases from bottom to top.
Here and in all other graphs all quantities are dimensionless.}
\end{figure}

In Fig. \ref{figQDLqcXXZ0} we show both $\mathcal{QD}$
and $\mathcal{L}_{QC}^{x}$ for the present model as a function of 
$\Delta$ for several values of $T$. 
Quantum discord is 
by its definition always bounded, $0\leq\mathcal{QD}\leq1$, while 
$\mathcal{L}_{QC}^{x}$ is unbounded, diverging at the QCPs 
($\Delta = \pm1$). This feature is 
clearly illustrated in Fig. \ref{figQDLqcXXZ0}.

\begin{figure}[!ht]
\includegraphics[width=8cm]{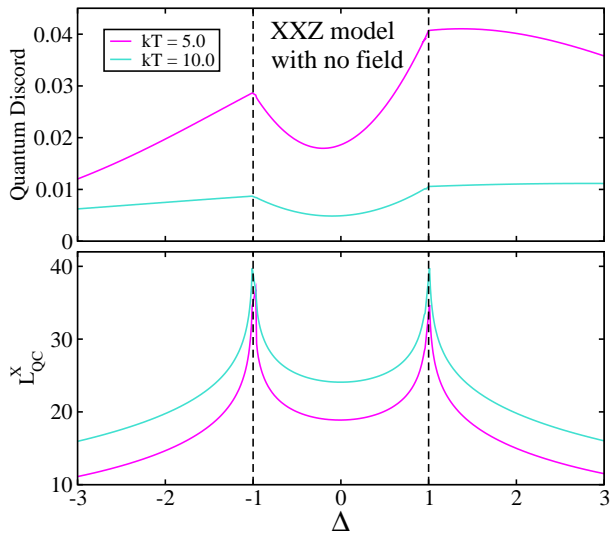}
\caption{\label{fig_ap1}(color online)  
Same as Fig. \ref{figQDLqcXXZ0}, but now we focus on higher values of
temperature.}
\end{figure}

Note also that the QD is able to pinpoint the correct location of the 
QCPs up to $kT=10.0$. This is clearly seen by looking at Fig. \ref{fig_ap1}, where we can better appreciate the cusps of 
$\mathcal{QD}$ at the 
two QCPs.
And, as expected for a reasonable QCP detector, as we increase $T$ its
efficacy decreases. This should be contrasted with the 
opposite behavior of $\mathcal{L}_{QC}^{x}$, being even ``sharper'' 
to detect a QCP for higher $T$. 
As it is clear now, this fact is due to its being ill-defined at the 
QCPs for this model.

\begin{figure}[!ht]
\includegraphics[width=8cm]
{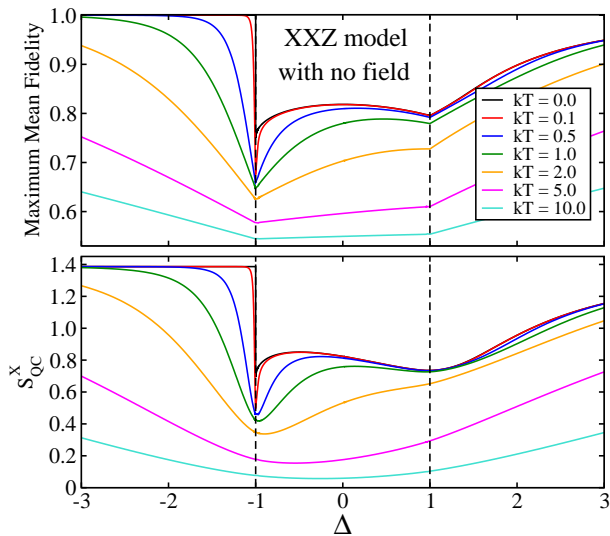}
\caption{\label{figFidSqcXXZ0}(color online)  
Maximum mean fidelity, Eq.~(\ref{fmax2a}), 
and the coherence entropy, 
Eq.~(\ref{sqc}), as a function of $\Delta$ for several values of 
temperature. In both panels the temperature increases from top
to bottom.}
\end{figure}

\begin{figure}[!ht]
\includegraphics[width=8cm]
{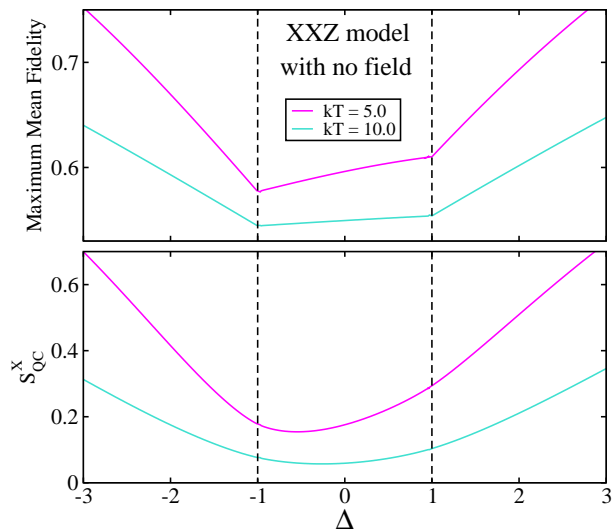}
\caption{\label{fig_ap2}(color online)  
Same as Fig. \ref{figFidSqcXXZ0}, but now we focus on 
higher values of temperature. Note that the 
first order derivatives of $\overline{\mathcal{F}}_{ext}$
are still discontinuous at both QCPs while for 
$\mathcal{S}_{QC}^x$ no clear indication for the QCPs can be seen.}
\end{figure}

In Fig. \ref{figFidSqcXXZ0} we plot $\overline{\mathcal{F}}_{ext}$ and
$\mathcal{S}_{QC}^{x}$. Both quantities now are bona fide QCP detectors.
For $T=0$ they are both discontinuous at $\Delta=-1$, while at finite
$T$ it is only $\overline{\mathcal{F}}_{ext}$ that has a discontinuous first order derivative at this QCP. This feature is also present for 
$\overline{\mathcal{F}}_{ext}$ at the second QCP, either at $T=0$ or
when $T>0$. On the other hand, $\mathcal{S}_{QC}^{x}$ is less sharp to 
pinpoint the second QCP when compared with its ability to 
detect the first one. Also, when we increase $T$, 
$\mathcal{S}_{QC}^{x}$ loses its ability to detect both QCPs before this happens with $\overline{\mathcal{F}}_{ext}$. 
This is clearly depicted in Fig. \ref{fig_ap2}.

The results presented for this model are very illustrative of a general
trend that will show up in the following ones. The most important message
so far is that $\mathcal{L}_{QC}^K$, the logarithm of the spectrum, 
should not be considered a reliable QCP detector. It is unbounded and 
ill-defined when at least one of the eigenvalues of the operator 
$[\rho_{23},K]^2$ is zero. This happens for this model exactly 
in the QCPs and it is a consequence of the symmetry of the Hamiltonian 
and not a particular ability of $\mathcal{L}_{QC}^K$ to detect
QCPs. Indeed, we showed that this feature, the existence of null eigenvalues, will persist even when $T\rightarrow \infty$. This means
that $\mathcal{L}_{QC}^K$ diverges at the two QCPs for this model no matter how high the temperature is, a clear indication that 
$\mathcal{L}_{QC}^K$ cannot be considered a useful or well-defined 
QCP detector. 

Second, the other quantities studied here are all bona fide QCP detectors.
They share key characteristics of all known QCP detectors, namely,
they are bounded and their efficacy to pinpoint a QCP diminishes 
as we increase the temperature. 
Two of these quantifies, the teleportation based QCP detector and the quantum discord, are more robust to temperature 
increases than the other one, the coherence entropy 
$\mathcal{S}_{QC}^K$. However, disregarding the issue of 
scalability for high spin systems and their operational meaning, 
from a strictly theoretical point of view they tend to complement each
other in the investigation of QCPs as we will show next for the other
models.

\subsection{The XXZ model in an external field}
\label{xxzh}

Using the same notation and conventions given in Sec. \ref{xxz0},
the Hamiltonian for the XXZ model in the presence of an external
longitudinal magnetic field is \cite{yan66,clo66,klu92,bor05,boo08,tri10,tak99}
\begin{equation}
H = \sum_{j=1}^{L}\left(\sigma^{x}_{j}\sigma^{x}_{j+1} +
\sigma^{y}_{j}\sigma^{y}_{j+1} + \Delta
\sigma^{z}_{j}\sigma^{z}_{j+1} - \frac{h}{2}\sigma^z_j\right), \label{Hxxzh}
\end{equation}
with $h$ denoting the external magnetic field.

For a finite magnetic field $h$, the ground state of this model 
has two QCPs \cite{yan66,clo66,klu92,bor05,boo08,tri10,tak99}. 
At the first one, $\Delta_1$, the ground state changes 
from a ferromagnetic ($\Delta < \Delta_1$) to a critical 
antiferromagnetic phase ($\Delta_1 < \Delta < \Delta_{2}$). 
At the second one, $\Delta_{2}$, it becomes  
an Ising-like antiferromagnet for $\Delta > \Delta_{2}$. 

The value of $\Delta_1$ is related to the external field by 
the following expression,
\begin{equation}
h=4J(1+\Delta_1),
\label{d1}
\end{equation}
while $\Delta_2$ is computed once we know $h$ by solving
the following equation,
\begin{equation}
h=4\sinh(\eta)\sum_{j=-\infty}^\infty\frac{(-1)^j}{\cosh(j\eta)},
\label{dinf}
\end{equation}
where $\eta = \cosh^{-1}(\Delta_{2})$.

In Table \ref{d1dinf} we give the solutions to Eqs.~(\ref{d1})
and (\ref{dinf}) for $h=12.0$, the external field we will
be using in this work. We should note, nevertheless, that the results here
reported are quite general, being valid for other 
fields too \cite{pav23,pav24}.  
For comparison, we also provide in Table \ref{d1dinf}
the two QCPs when we have no field (the model we studied in 
the previous section). 
\begin{table}[!ht]
\caption{\label{d1dinf}
Quantum critical points $\Delta_1$ and $\Delta_2$  
for the case of no field and when the external field is $h=12.0$. 
When $h=12.0$, $\Delta_{2}$ listed below is accurate within a 
numerical error of $\pm 0.001$.} 
\begin{ruledtabular}
\begin{tabular}{lrr}
& $h=0$ & $h=12$\\ \hline 
$\Delta_1$ & -1.00 & 2.00\\ 
$\Delta_{2}$  & 1.00 & 4.875\\
\end{tabular}
\end{ruledtabular}
\end{table} 

When we turn on the longitudinal field, 
the magnetization $z$ is no longer null but we 
still have $xx=yy$. In this case, the four eigenvalues used to
compute $\mathcal{S}_{QC}^{x}$ and $\mathcal{L}_{QC}^{x}$ are
\begin{eqnarray}
\alpha_{1,2}^x&=&-\frac{1}{4}\left(zz + \sqrt{xx^2+z^2}\right)^2, 
\label{a12XXZh} \\
\alpha_{3,4}^x&=&-\frac{1}{4}\left(zz - \sqrt{xx^2+z^2}\right)^2. 
\label{a34XXZh}
\end{eqnarray}
Similar to the case with no field, we still have that 
$\mathcal{S}_{QC}^{x}=\mathcal{S}_{QC}^{y}$ and $\mathcal{L}_{QC}^{x} =\mathcal{L}_{QC}^{y}$. This happens because 
the eigenvalues defining those quantities are all equal. We also have
that the eigenvalues, Eqs.~(\ref{a12XXZh}) and (\ref{a34XXZh}), are all negative. 

\begin{figure}[!ht]
\includegraphics[width=8cm]
{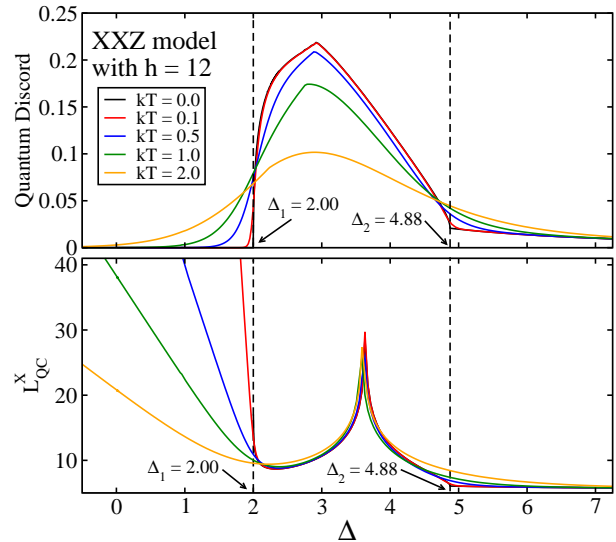}
\caption{\label{figQDLqcXXZh}(color online)  
Quantum discord, Eq.~(\ref{QD}), 
and the logarithm of the spectrum, 
Eq.~(\ref{lqc}), as a function of $\Delta$ for several values of 
temperature when $h=12.0$. 
For the upper panel, the temperature increases from top
to bottom between the QCPs, and for the lower panel, it increases from top to bottom when $\Delta < \Delta_1$.}
\end{figure}

In Fig. \ref{figQDLqcXXZh} we show $\mathcal{QD}$ and 
$\mathcal{L}_{QC}^x$ assuming a field of $h=12.0$.
The problematic definition of $\mathcal{L}_{QC}^x$ now manifests 
itself in two different places. First, for $\Delta < \Delta_1$, 
numerical analysis shows that $\alpha_{3,4}^x$ tends monotonically
to zero as $\Delta\rightarrow -\infty$. Moreover, the lower
$T$ the faster $\alpha_{3,4}^x$ approaches zero. 
This is why we see those extremely high values for 
$\mathcal{L}_{QC}^x$ when $\Delta < \Delta_1$. Second, between
the two QCPs there is a value of $\Delta$ such that  
$\alpha_{1,2}^x\rightarrow 0$ and thus $\mathcal{L}_{QC}^x$ becomes 
undefined at this point. 
This is the reason for the cusps between the 
two QCPs that is not associated to any QPT seen in all curves for
$\mathcal{L}_{QC}^x$. 

The curves for QD also has a cusp between the two QCPs that are not 
related to a QPT. This is related to the minimization procedure of 
the quantum conditional entropy appearing in its definition 
\cite{wer10,wer10b}. In the present notation, it is related to a 
discontinuous change in the optimal value of $\theta$ that minimizes Eq.~(\ref{QD}). However, as we increase $T$ this cusp smooths out and 
disappear while for $\mathcal{L}_{QC}^x$ the cusp is always there.

At $T=0$, the curve for the QD as a function of $\Delta$ has discontinuous derivatives exactly at the two QCPs. These cusps are smoothed out and
displaced away from the correct location of the QCPs as we increase 
$T$. The curves for $\mathcal{L}_{QC}^x$ behave similarly in the
vicinity of the QCPs. 

\begin{figure}[!ht]
\includegraphics[width=8cm]
{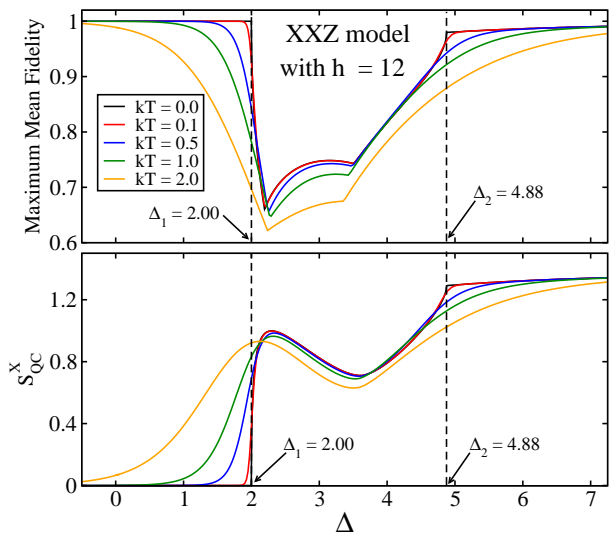}
\caption{\label{figFidSqcXXZh}(color online)  
Maximum mean fidelity, Eq.~(\ref{fmax2a}), 
and the coherence entropy, 
Eq.~(\ref{sqc}), as a function of $\Delta$ for several values of 
temperature when $h=12.0$. 
For the upper panel, the temperature increases from top
to bottom when $\Delta < \Delta_1$. For the lower panel, 
it increases from bottom to top when $\Delta < \Delta_1$.}
\end{figure}

In Fig. \ref{figFidSqcXXZh} we show $\overline{\mathcal{F}}_{ext}$ 
and $\mathcal{S}_{QC}^x$ when $h=12.0$ for several values of $T$.
Looking at the curves of $\overline{\mathcal{F}}_{ext}$ and
$\mathcal{S}_{QC}^x$ when $T=0$, we note that the QCPs are all detected
by discontinuities in the derivatives of those quantities as a function
of $\Delta$. As we increase $T$, those discontinuous derivatives (cusps)
are smoothed out and displaced from the correct spot of the QCPs. We
should also note that $\overline{\mathcal{F}}_{ext}$ has two 
extra cusps between the two QCPs that are related to
the maximization over the sets $S_k$ of unitary operations 
available to Bob \cite{pav23,pav24}. These extras cusps are not related to 
QPTs and they are located around the local maxima and minima seen in the 
curves for $\mathcal{S}_{QC}^x$ between the two QCPs. 

\begin{figure}[!ht]
\includegraphics[width=8cm]
{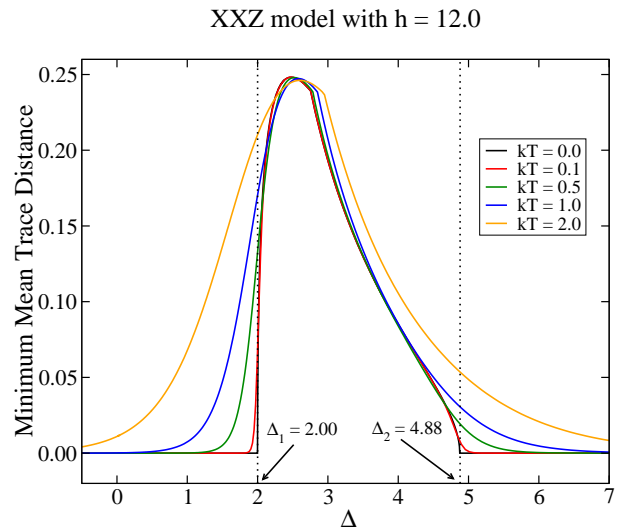}
\caption{\label{figtrDXXZh}(color online)  
Minimum mean trace distance, Eq.~(\ref{dmin2}), 
as a function of $\Delta$ for several values of 
temperature when $h=12.0$. The temperature increases from
bottom to top when $\Delta < \Delta_1$.}
\end{figure}

In Fig. \ref{figtrDXXZh} we show $\overline{\mathcal{D}}_{int}$,
the minimum mean trace distance for the internal teleportation protocol,
as a function of $\Delta$. We have fixed the field at $h=12.0$ and plotted
$\overline{\mathcal{D}}_{int}$ for several values of $T$.

In Fig. \ref{figtrDXXZh} we see that for $T=0$ the QCPs are detected by discontinuities in the derivatives of $\overline{\mathcal{D}}_{int}$ 
with respect to $\Delta$ (see the kinks at $\Delta_1$ and $\Delta_2$).  
Contrary to $\overline{\mathcal{F}}_{ext}$, 
we now have only one tiny kink between the two QCPs that is not 
associated with a QPT. Similarly to the origin
of the two extra kinks of $\overline{\mathcal{F}}_{ext}$, the single 
extra kink of $\overline{\mathcal{D}}_{int}$ 
can be traced back to the minimization over the sets $S_k$ 
of unitary operations \cite{pav24}.  When $T > 0$,
the kinks related to the two QCPs are smoothed out and displaced
from the exact locations of the QCPs. The extra kink between the
two QCPs is also displaced from its $T=0$ spot but 
not appreciably smoothed
out in the ranges of temperatures shown in Fig. \ref{figtrDXXZh}.

In order to be more quantitative, we now compare the efficacy of the above
quantities in estimating the correct values for the QCPs with finite $T$
data alone. We adopt the same techniques fully described in Refs. \cite{wer10b,pav23,pav24} in the following analysis. 

Although for finite $T$ the kinks are smoothed out,
we still have an abrupt change in the 
value of those quantities about the QCPs. As such, for a fixed $T$, 
we compute the derivatives of those curves with respect to $\Delta$ 
about the QCPs. Then we pick the value of $\Delta$ giving the greatest magnitude for the derivatives. This $\Delta$ is considered the best 
approximation to the value of the QCP at that fixed $T$. Repeating this 
procedure for several temperatures, we can extrapolate to $T=0$ and 
correctly arrive at the exact values for the QCPs. 

\begin{figure}[!ht]
\includegraphics[width=8cm]
{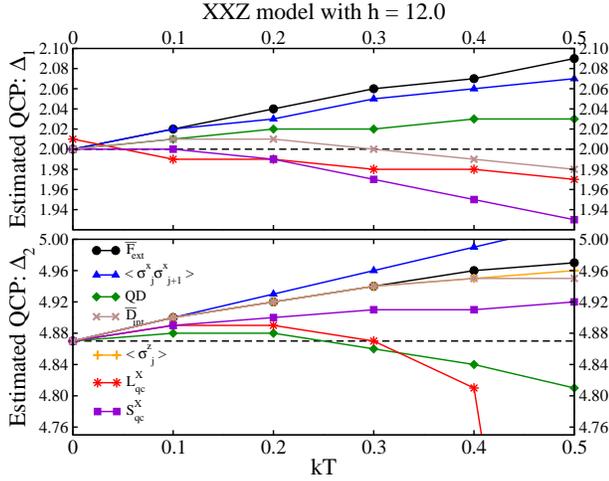}
\caption{\label{fig_estimativa1}(color online)  
Estimated QCPs using finite $T$ data according to the procedure 
explained in the main text. For each value of $kT$,
and for $f=\mathcal{QD},\mathcal{S}_{QC}^x,\mathcal{L}_{QC}^x, \overline{\mathcal{F}}_{ext},\overline{\mathcal{D}}_{int},$ 
$ \langle \sigma^x_j\sigma^x_{j+1} \rangle, 
\langle \sigma^z_j\rangle$ (see legend),
the upper panel gives the values of $\Delta$ related to 
the extrema of $df/d\Delta$ and 
the lower panel the values of $\Delta$ associated with the extrema 
of $d^2f/d\Delta^2$.
The dashed lines mark the exact values of the QCPs. 
See text for details.}
\end{figure}

We work with six different temperatures, i.e., 
$kT=0, 0.1, 0.2, 0.3, 0.4, 0.5$. For each one of these temperatures,
we compute as a function of $\Delta$ and in increments of $0.01$
the several quantities shown in Fig. \ref{fig_estimativa1}. Then, 
about $\Delta_1$, we numerically evaluate the first order derivatives 
of those quantities, picking the value of $\Delta$ that gives the 
greatest magnitude for the derivatives. About $\Delta_2$, we numerically compute the second order derivatives of those quantities, picking again 
the value of $\Delta$ giving the 
greatest magnitude for the second order derivatives. 
The values of those $\Delta$'s are plotted in Fig. \ref{fig_estimativa1}. 


In numerically computing those derivatives, we used the forward
difference method, namely, $df(x)/dx \approx [f(x+\eta)-f(x)]/\eta$,
where $\eta=0.01$. Also, since $\Delta$ was changed in 
steps of $0.01$, the spot of the maxima of the absolute value of 
the first order derivatives have a numerical error of $\pm 0.01$. 
For the second order derivatives, which are computed from the first order ones, we have that the errors related to the location of their extrema 
are at least $\pm 0.02$ \cite{pav23,pav24}. 

Looking at the upper panel of Fig. \ref{fig_estimativa1}, 
we realize that the internal teleportation based QCP detector
($\overline{\mathcal{D}}_{in}$)
outperforms all quantities in estimating the QCP $\Delta_1$ 
when $kT\geq 0.2$. For $kT\leq0.2$, and taking into account the 
numerical error ($\pm 0.01$) for the values of the estimated QCP,  
we have that the QD, 
$\overline{\mathcal{D}}_{in}$, $\mathcal{S}_{QC}^x$, and  
$\mathcal{L}_{QC}^x$ are all equivalent in estimating the correct value
of $\Delta_1$. Moreover, all quantities tend to
the correct value of the QCP more or less linearly below a certain 
value of $kT$. 

Moving our attention to the 
lower panel of Fig. \ref{fig_estimativa1}, 
we note that the quantum discord ($\mathcal{QD}$) stands out 
as the optimal choice to estimate the correct value of $\Delta_2$, the second QCP. This is true for all the temperature range shown in 
Fig. \ref{fig_estimativa1}. Taking into account that 
now the numerical error is at least $\pm 0.02$ for the estimated 
value of $\Delta_2$, 
below $kT \approx 0.1$ we have that the QD is equivalent to
$\mathcal{S}_{QC}^x$ and $\mathcal{L}_{QC}^x$ in estimating the second
QCP. We should also note that $\mathcal{L}_{QC}^x$ is almost useless to
estimate the second QCP for $kT \geq 0.4$, providing very poor estimates.

Before we move on, we should note that this case, 
which was not studied in Ref. \cite{li20}, 
clearly illustrates that $\mathcal{L}_{QC}^x$ 
is not the optimal QCP detector. For this model, the internal teleportation based QCP detector is the best choice to estimate 
$\Delta_1$ while the QD is the best choice for estimating $\Delta_2$
when only finite $T$ data are available.

\subsection{The XY model}
\label{xy}

Following the notation and boundary conditions already
explained for the previous models, 
the Hamiltonian for the one-dimensional XY model subjected to a 
transverse magnetic field is \cite{lie61,bar70,bar71},
\begin{equation}
H \!=\! -\frac{\lambda}{4}\!\sum_{j=1}^{L}\!\left[(1+\gamma)\sigma^{x}_{j}\sigma^{x}_{j+1} + (1-\gamma)\sigma^{y}_{j}\sigma^{y}_{j+1}\right]\! 
- \frac{1}{2}\!\sum_{j=1}^{L}\!\sigma^z_j. \label{Hxy}
\end{equation}
Here $\lambda$ is related to the inverse of the  
external magnetic field strength and
$\gamma$ is the anisotropy parameter. When $\gamma=\pm 1$ we obtain the 
transverse Ising model and when $\gamma=0$ we get the isotropic 
XX model in a transverse field.

If we fix $\gamma$ and change $\lambda$, i.e., if we change the external
field, we have a QCP at $\lambda_c=1.0$.  This is 
the Ising transition. For $\lambda < 1$ the ground state is an ordered ferromagnet while for $\lambda > 1$ it becomes a quantum paramagnet  \cite{pfe70}. If we now fix $\lambda$ such that $\lambda > 1$, we also have another QPT if we change $\gamma$. It is the anisotropy transition,
whose QCP is located at $\gamma_c=0$ \cite{lie61,bar70,bar71,zho10}.
In this case, one of the phases is an ordered ferromagnet
in the x-direction while the other phase is an ordered ferromagnet 
in the y-direction. These two QPTs belong to different universality classes \cite{lie61,bar70,bar71,zho10}.

For arbitrary values of $\lambda$ and $\gamma$, the density matrix 
describing a pair of spins for this model is given by Eqs.~(\ref{rho23})
and (\ref{varA})-(\ref{varE}). In general we have all two-point 
correlation functions different ($xx\neq yy\neq zz$) and a non-null
magnetization ($z\neq 0$). Therefore, using 
Eqs.~(\ref{varA})-(\ref{varE}), we can write Eqs.~(\ref{a1z}) and
(\ref{a3z}), the eigenvalues defining
$\mathcal{S}_{QC}^z$ and $\mathcal{L}_{QC}^z$, as follows,
\begin{eqnarray}
\alpha_{1,2}^z&=&-(xx + yy)^2/4, \label{a12XY} \\
\alpha_{3,4}^z&=&-(xx - yy)^2/4. \label{a34XY}
\end{eqnarray}
Note that the authors of Ref. \cite{li20} did not work with 
$\mathcal{S}_{QC}^{x,y}$ and $\mathcal{L}_{QC}^{x,y}$ for this model.

Again, all eigenvalues are negative and for non zero two-point 
correlation functions, two of them become zero if $|xx|=|yy|$. If
we look at Eqs.~(\ref{a12XXZ0}) and (\ref{a34XXZ0}), we realize that
$\alpha_{1,2}^z$ and $\alpha_{3,4}^z$ have the same functional form of
$\alpha_{1,2}^x$ and $\alpha_{3,4}^x$, the eigenvalues
related to the XXZ model with no field. Indeed, changing $yy$ by $zz$ in the above expressions for the eigenvalues leads to the ones given by Eqs.~(\ref{a12XXZ0}) and
(\ref{a34XXZ0}). Therefore, the analysis we have made in 
Sec. \ref{xxz0} about certain symmetries of the Hamiltonian and
the equality of a pair of two-point correlation functions for any
temperature applies here. As we show next, the alleged ``robustness''
of $\mathcal{L}_{QC}^{z}$ in detecting the $\gamma$-transition is 
associated with a particular symmetry of the Hamiltonian and to the
ill-defined character of $\mathcal{L}_{QC}^{z}$. In other words,
once more this example will illustrate that there is nothing 
outstanding in $\mathcal{L}_{QC}^{z}$ that sets it apart from other 
QCP detectors. 

\subsection{The $\gamma$ transition}

The anisotropy transition occurs at $\gamma_c=0$. At this QCP, the 
Hamiltonian (\ref{Hxy}) becomes
\begin{equation}
H =-\frac{\lambda}{4}\!\sum_{j=1}^{L}\!\left[\sigma^{x}_{j}\sigma^{x}_{j+1} + \sigma^{y}_{j}\sigma^{y}_{j+1}\right]\! 
- \frac{1}{2}\!\sum_{j=1}^{L}\!\sigma^z_j. \label{Hxy0}
\end{equation}
Looking at Eq.~(\ref{Hxy0}), it is obvious that 
the two-point correlation functions $xx$ and $yy$ are equal. This 
happens because of the invariance of the Hamiltonian for 
rotations around the $z$-axis when $\gamma=0$. 

Since $xx=yy$ for any $T$, we immediately
see from Eq.~(\ref{a34XY}) that $\alpha_{3,4}^z=0$ for any value of 
temperature. In other words, $\mathcal{L}_{QC}^{z}$ will diverge 
at $\gamma=0$ no matter how high the temperature is. This is
the reason for the ``robustness'' of $\mathcal{L}_{QC}^{z}$ to 
temperature increases reported in Ref. \cite{li20}. As we understand now,
this robustness is misleading. It is simply a consequence of the 
ill-defined character of $\mathcal{L}_{QC}^{K}$ when any one of the eigenvalues appearing in its definition becomes zero.

\begin{figure}[!ht]
\includegraphics[width=8cm]
{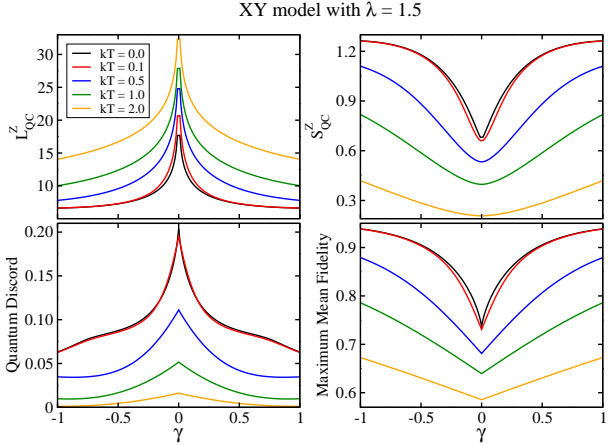}
\caption{\label{fig_lambda1p5}(color online)  
The several QCP detectors investigated in this work as a function of
$\gamma$. We fixed $\lambda=1.5$ in Hamiltonian (\ref{Hxy}).
In the top-left panel, temperature increases from bottom to top. In
the remaining panels, temperature increases from top to bottom. See 
text for details.}
\end{figure}

In Fig. \ref{fig_lambda1p5} we show for several values of temperature 
$\mathcal{L}_{QC}^{z}$ and the other
relevant QCP detectors, namely, $\mathcal{S}_{QC}^z, \mathcal{QD},$
and $\overline{\mathcal{F}}_{ext}$, 
as functions of $\gamma$. We fix 
$\lambda=1.5$ in all those curves.

Looking at Fig. \ref{fig_lambda1p5}, we realize that all quantities 
are useful in spotlighting the QCP up to $kT=2.0$. The QD and 
the external teleportation QCP detector (lower panels) spotlight 
the QCP by discontinuities in their first order derivatives with 
respect to $\gamma$ at the exact location of the QCP (see the kinks
at $\gamma_c=0$). Those
kinks are not displaced as we increase $T$ but become smoother.
On the other hand, $\mathcal{S}_{QC}^z$ does not have those kinks.
The QCP in this case is detected by noting that the 
global minimum of
$\mathcal{S}_{QC}^z$ occurs at $\gamma_c=0$. Finally,
$\mathcal{L}_{QC}^z$ diverges at the location of the QCP for any $T$
due to the reasons given above.

\subsection{The $\lambda$ transition}
\label{lambda}

The only case studied in Ref. \cite{li20} related to the $\lambda$ 
transition was the Ising transverse model ($\gamma=\pm1.0$). 
For definiteness, here we fix $\gamma=1.0$ in the following analysis.

\begin{figure}[!ht]
\includegraphics[width=8cm]
{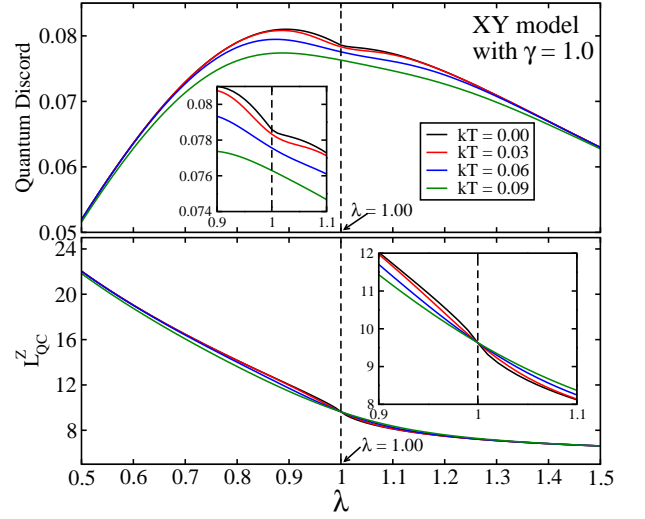}
\caption{\label{fig_gamma1p0}(color online)  
Quantum discord, Eq.~(\ref{QD}), 
and the logarithm of the spectrum, 
Eq.~(\ref{lqc}), as a function of $\lambda$ for several values of 
temperature. Here we fix $\gamma=1.0$ (Ising transverse model).
In all panels, the temperature increases from top
to bottom when $\lambda<1.0$. The insets zoom in at the QCP, where
in the $x$-axis we have $0.9\leq \lambda \leq 1.1$.}
\end{figure}

\begin{figure}[!ht]
\includegraphics[width=8cm]
{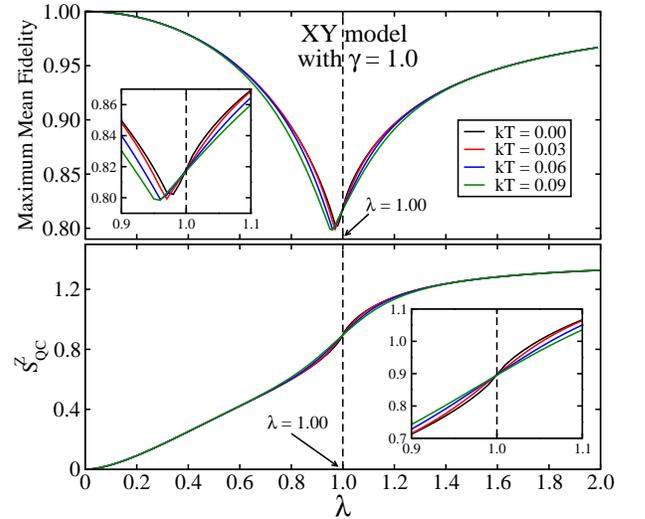}
\caption{\label{fig_gamma1p0b}(color online)  
Maximum mean fidelity, Eq.~(\ref{fmax2a}), 
and the coherence entropy, 
Eq.~(\ref{sqc}), as a function of $\lambda$ for several values of 
temperature. Here we fix $\gamma=1.0$. In both panels the temperature increases from top to bottom after $\lambda > 1.0$.}
\end{figure}

In Figs. \ref{fig_gamma1p0} and \ref{fig_gamma1p0b} we plot the QD, 
$\mathcal{L}_{QC}^z$, $\overline{\mathcal{F}}_{ext}$ 
and $\mathcal{S}_{QC}^z$ as functions of $\lambda$ for several values 
of $T$. At $T=0$, the QCP is given by inflection points of these quantities at $\lambda_c=1.0$. As we increase $T$, the inflection points move away from the QCP and become less prominent as we increase $T$.

\begin{figure}[!ht]
\includegraphics[width=8cm]
{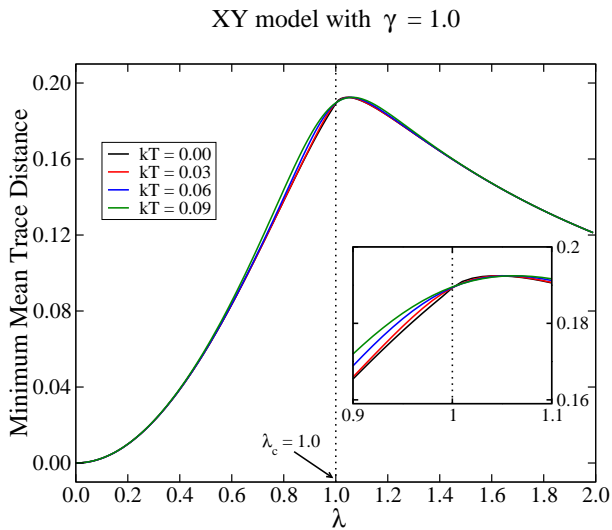}
\caption{\label{fig_gamma1p0c}(color online)  
Minimum mean trace distance, Eq.~(\ref{dmin2}), 
as a function of $\lambda$ for several values of 
temperature when $\gamma=1.0$. 
The temperature increases from bottom to top when
$\lambda < 1.0$. The inset zooms in at the QCP, where
in the $x$-axis we have $0.9\leq \lambda \leq 1.1$.}
\end{figure} 

Figure \ref{fig_gamma1p0c} illustrates the behavior of 
$\overline{\mathcal{D}}_{int}$ as a function of $\lambda$ for 
several values of $T$ and with $\gamma=1.0$. At $T=0$, it is now $d\overline{\mathcal{D}}_{int}/d\lambda$ that has an inflection point
at the QCP \cite{pav24}. Similarly to the above cases, the higher 
$T$ the more distant from the QCP the inflection point is and it also
becomes less prominent as we increase $T$.

\begin{figure}[!ht]
\includegraphics[width=8cm]
{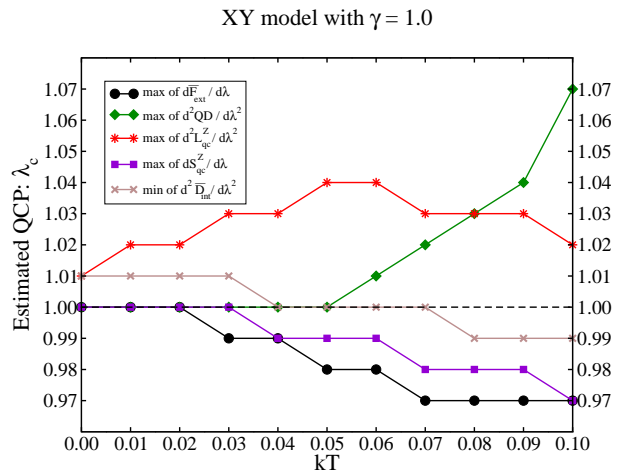}
\caption{\label{fig_gamma1p0d}(color online)  
Estimated QCPs using finite $T$ data according to the procedure 
given in the main text. For each value of $kT$, 
we plot the values of $\lambda$ yielding 
the maximum of $df/d\lambda$, where 
$f=\overline{\mathcal{F}}_{ext}, \mathcal{S}_{QC}^z$, and
the values of $\lambda$ giving the maximum 
of $d^2f/d\lambda^2$, where 
$f=\mathcal{QD},\overline{\mathcal{D}}_{int},\mathcal{L}_{QC}^z$.
The dashed line marks the exact location of the QCP. 
See text for details.}
\end{figure}

To quantitatively compare the performance of these QCP detectors 
when only finite $T$ data are available, 
we repeat here the same analysis carried
out for the XXZ model in an external field (see Sec. \ref{xxzh}). We now 
work with eleven different temperatures. 
We start at $kT=$ $0$ and 
go up to $kT=0.1$ in increments of $0.01$. For each value of $kT$,
we compute the first and second order derivatives 
with respect to $\lambda$ for all the five QCP detectors shown in 
Figs. \ref{fig_gamma1p0}-\ref{fig_gamma1p0c}. 
In Fig. \ref{fig_gamma1p0d} we only show the corresponding 
estimate for the QCP extracted from the derivative
(first or second order) leading to the best performance for each one of
those quantities. As in Sec. \ref{xxzh}, the critical point is estimated
by picking the value of $\lambda$ giving the greatest magnitude of the 
respective derivative around the exact $T=0$ location of the QCP.

Contrary to what we did in Sec. \ref{xxzh}, we numerically computed 
those derivatives using the central difference method, namely, 
$df(x)/dx \approx [f(x+\eta)-f(x-\eta)]/(2\eta)$,
where $\eta=0.01$.  Both methods are equally valid but we opted now 
to work with the central difference method to contrast it with the 
curve of Ref. \cite{pav24}, where the forward difference method
was used to compute $\overline{\mathcal{D}}_{int}$. 
Since $\lambda$ was changed in 
steps of $0.01$, both methods will differ by at least $0.01$ in 
giving the maximum for the magnitude of the derivatives. This is an
illustrative example showing that the results reported here and 
in Ref. \cite{pav23,pav24} are accurate by at most the step
$\eta=0.01$ used to generate the curves (see Appendix for more details).
Furthermore, the same numerical errors related to the first and 
second order derivatives reported in Sec. \ref{xxzh} apply here. 
The spot of the maxima of the absolute value of 
the first and second order derivatives have a numerical error of 
$\pm 0.01$ and $\pm 0.02$, respectively \cite{pav23,pav24}. 

Looking at Fig. \ref{fig_gamma1p0d}, we note that
within the numerical errors (up to $\pm 0.02$ for the second derivatives),
all quantities but $\mathcal{L}_{QC}^z$ give the correct location of 
the QCP for $kT\leq 0.4$. Moreover, within an error of $\pm0.01$, 
the internal teleportation based QCP detector 
($\overline{\mathcal{D}}_{int}$) gives the correct value of the QCP for 
all temperatures shown in Fig. \ref{fig_gamma1p0d}. Again, the bottom
line is that neither $\mathcal{S}_{QC}^z$ nor $\mathcal{L}_{QC}^z$ stands
out as the most efficient QCP detector when only finite $T$ data are 
at stake.

\section{Conclusion}
\label{conclusion}

We analyzed qualitatively and quantitatively the efficacy 
of different tools to detect quantum critical points (QCPs)
for several classes of quantum phase transitions.
In particular, we focused on the efficacy of those tools to 
properly identify a QCP when we only have access to finite temperature data. We worked with the following quantities, namely, 
the quantum discord \cite{oll01,hen01,wer10,wer10b}, 
the coherence entropy and the logarithm of the spectrum \cite{li20}, 
which are based on the 
the quantum coherence \cite{wig63,fan14,gir14}, 
and the external and internal teleportation based QCP detectors \cite{pav23,pav24},
built on top of the standard quantum teleportation protocol \cite{ben93,yeo02,rig17}.

We worked with several types of one dimensional spin-1/2 chains,
whose ground states had at least two quantum phases. The 
spin chains were studied in the thermodynamic  
limit (infinite number of spins) and we assumed the spin chains to be
in equilibrium with a heat bath at temperature $T$. 
The models we studied were the XXZ model with and without an external  
longitudinal magnetic field, the Ising transverse model, 
and the XY model in an external transverse magnetic field. 
 
From the theoretical point of view, 
the most important ingredient needed to compute all the quantities 
above is the density matrix describing a pair of nearest neighbor
spins from the chain. It can be completely determined once we 
know all the one- and two-point correlation functions for a
given spin chain. After obtaining 
this density matrix as a function of the
temperature and of the parameters of the system's Hamiltonian, we
were able to compute the above quantities in the vicinity of the 
QCPs for several different values of $T$. This allowed was to
investigate quantitatively the accuracy of those QCP detectors in
correctly spotlighting the location of a QCP using finite $T$ data.

The first major result we arrived at is related to the faulty 
definition of the logarithm of the spectrum 
$\mathcal{L}_{QC}^K$ \cite{li20}. 
We showed that it cannot be defined for several classes of two-qubit
density matrices. Also, we showed that the alleged ``robustness'' 
of $\mathcal{L}_{QC}^K$ in spotlighting QCPs at finite $T$ is related
to this ill-definedness and has nothing to do with an intrinsic 
robustness to detect QCPs. Indeed, we showed that for all the models investigated in Ref. \cite{li20}, with the exception of the Ising model, the locations of the QCPs are exactly 
where $\mathcal{L}_{QC}^K$ is ill-defined (it diverges). This is 
the underlying cause for its ``robustness'' in detecting a QCP. We
also showed that $\mathcal{L}_{QC}^K$ may give us false alarms, 
being divergent (undefined) when no quantum phase transition is taking
place.

The second major result is related to the fact that no single QCP detector
outperforms all the others. The optimal QCP detector depends on the model, on the QCP, and on the temperature. However, almost always either 
the quantum discord or the teleportation based QCP detector
is the optimal choice. For the Ising transverse model, though, when 
we have low temperatures the
coherence entropy $\mathcal{S}_{QC}^K$ is as efficient as 
the quantum discord and the internal teleportation based QCP detector.

We end by noting that the comparative study among 
the QCP detectors here presented does not consider the computational complexity to 
calculate them, in particular for high spin systems, 
as well as their operational meaning and experimental feasibility. 
It is known that the quantum discord is not easily computed for 
high spin systems \cite{mal16}, being an NP-complete 
problem \cite{hua14}. This means that it becomes impracticable 
to compute the quantum discord as we increase the size of the system 
under investigation. Furthermore, the quantum discord does not have
a direct operational meaning and no general procedure to 
its direct measurement is known. We also have that
$\mathcal{L}_{QC}^K$ and $\mathcal{S}_{QC}^K$ \cite{li20}  
are not easily computed for high spins and that no direct way of 
measuring them is available \cite{pav23}. On the other hand,
the teleportation based QCP detectors have a direct experimental 
meaning and are easily generalized to high spin systems \cite{pav23,pav24}. On top of that, 
the necessary experimental steps to the
implementation of the teleportation based QCP detectors 
are already at hand \cite{ron15,bra19,xie19,noi22,xue22,mad22,xie22,gam17,ohn99,han03,zin10,pet22}.

\begin{acknowledgments}
GR thanks the Brazilian agency CNPq
(National Council for Scientific and Technological Development) for funding and 
CNPq/FAPERJ (State of Rio de Janeiro Research Foundation) for financial support
through the National Institute of Science and Technology for Quantum Information.
GAPR thanks the São Paulo Research Foundation (FAPESP)
for financial support through the grant 2023/03947-0.

\end{acknowledgments}

\appendix*

\section{Forward and central difference methods}

The forward difference method to numerically compute the derivative
of a function at point $x$ is given by the following expression,
\begin{equation}
\frac{df(x)}{dx} \approx \frac{f(x+\eta)-f(x)}{\eta},
\end{equation}
where $\eta$ is the numerical step used to generate 
point $x_{j+1}$ from point $x_j$. 
The central difference method, 
on the other hand, is given by
\begin{equation}
\frac{df(x)}{dx} \approx \frac{f(x+\eta)-f(x-\eta)}{2\eta}.
\end{equation}

\begin{figure}[!ht]
\includegraphics[width=8cm]
{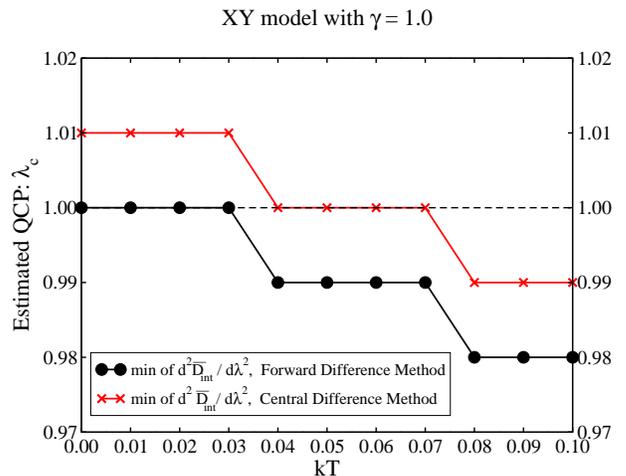}
\caption{\label{fig_apendice}(color online)  
Estimated QCPs using finite $T$ data according to the procedure 
given in the main text. The two curves are obtained from two different 
finite difference methods to approximate the derivatives of 
$\overline{\mathcal{D}}_{int}$. The dashed line marks the exact value of
the QCP ($\lambda_c=1.0$).}
\end{figure}

Applying twice either the forward method or the central method, 
we computed
for each value of $kT$ in Fig. \ref{fig_apendice} 
the second derivative of $\overline{\mathcal{D}}_{int}$ with respect 
to $\lambda$, the driving term for the Ising transverse model of 
Sec. \ref{lambda}. Then, we picked the value of $\lambda$ leading to 
the minimum of $d^2\overline{\mathcal{D}}_{int}/d\lambda^2$. The points
in Fig. \ref{fig_apendice} are the $\lambda$'s giving the minima 
of $d^2\overline{\mathcal{D}}_{int}/d\lambda^2$ at each $kT$.

Looking at Fig. \ref{fig_apendice}, we see that for every value of
$kT$ the location of 
the minimum of $d^2\overline{\mathcal{D}}_{int}/d\lambda^2$, obtained 
by the central difference method, differs by $0.01$ from the 
location of the minimum determined via the other method. En passant, we should mention that if we had computed the derivatives using the backward difference method, where 
$df(x)/dx \approx [f(x)-f(x-\eta)]/\eta$, we would have obtained a 
curve that in Fig. \ref{fig_apendice} would appear displaced above 
the one for the central difference method by $\eta=0.01$. 
This will always happen 
because the derivative computed by the central difference method
is the average of the values obtained by the forward and 
backward difference methods.



\begin{thebibliography}{200}

\bibitem{ost02} A. Osterloh, L. Amico, G. Falci, and R. Fazio, 
Nature (London) \textbf{416}, 608 (2002).

\bibitem{osb02} T. J. Osborne and M. A. Nielsen, Phys. Rev. A \textbf{66}, 
032110 (2002).

\bibitem{wu04}  L.-A. Wu, M. S. Sarandy, and D. A. Lidar, Phys. Rev. Lett. \textbf{93}, 250404 (2004).

\bibitem{ver04} F. Verstraete, M. Popp, and J. I. Cirac,
Phys. Rev. Lett. \textbf{92}, 027901 (2004).

\bibitem{oli06} T. R. de Oliveira, G. Rigolin, M. C. de Oliveira, and E. Miranda, 
Phys. Rev. Lett. \textbf{97}, 170401 (2006); 
T. R. de Oliveira, G. Rigolin, and M. C. de Oliveira, Phys. Rev. A 73, 010305 (2006); 
T. R. de Oliveira, G. Rigolin, M. C. de Oliveira, and E. Miranda, Phys. Rev. A \textbf{77}, 032325(R) (2008).

\bibitem{dil08} R. Dillenschneider, Phys. Rev. B \textbf{78}, 224413 (2008). 

\bibitem{sar08} M. S. Sarandy, Phys. Rev. A \textbf{80}, 022108 (2009).

\bibitem{fan14} G. Karpat, B. \c{C}akmak, and F. F. Fanchini,
Phys. Rev. B \textbf{90}, 104431 (2014).

\bibitem{gir14} D. Girolami, Phys. Rev. Lett. \textbf{113}, 
170401 (2014).

\bibitem{hot08} M. Hotta, Phys. Lett. A \textbf{372}, 5671 (2008).

\bibitem{hot09} M. Hotta, J. Phys. Soc. Jpn. \textbf{78}, 034001 (2009).

\bibitem{ike23a} K. Ikeda, Phys. Rev. D \textbf{107}, L071502 (2023).

\bibitem{ike23b} K. Ikeda, Phys. Rev. Appl. \textbf{20}, 024051 (2023).

\bibitem{ike23c} K. Ikeda, AVS Quantum Sci. \textbf{5}, 035002 (2023).

\bibitem{ike23d} K. Ikeda, R. Singh, and R.-J. Slager,  e-print arXiv: 2310.15936 [quant-ph].

\bibitem{sac99} S. Sachdev, \textit{Quantum Phase Transitions}
(Cambridge University Press, Cambridge, 1999).

\bibitem{gre02} M. Greiner, O. Mandel, T. Esslinger, T. W. H\"ansch, and I. Bloch, Nature (London) \textbf{415}, 39 (2002).

\bibitem{gan10} V. F. Gantmakher and V. T. Dolgopolov, Phys. Usp. 53, \textbf{1} (2010).

\bibitem{row10} S. Rowley, R. Smith, M. Dean, L. Spalek, M. Sutherland, M. Saxena, P. Alireza, C. Ko, C. Liu, E. Pugh \textit{et al.}, 
Phys. Status Solidi B \textbf{247}, 469 (2010).

\bibitem{woo98} W. K. Wootters, Phys. Rev. Lett. 80, 2245 (1998).

\bibitem{wer10} T. Werlang and G. Rigolin, Phys. Rev. A \textbf{81}, 044101 (2010).

\bibitem{wer10b} T. Werlang, C. Trippe, G. A. P. Ribeiro, and G. Rigolin, 
Phys. Rev. Lett. \textbf{105}, 095702 (2010); T. Werlang, G. A. P. Ribeiro, and G. Rigolin, Phys. Rev. A \textbf{83}, 062334 (2011); T. Werlang, G. A. P. Ribeiro, and G. Rigolin, Int.J. Mod. Phys. B \textbf{27}, 1345032 (2013).

\bibitem{li20} Y. C. Li, J. Zhang, and H.-Q. Lin, 
Phys. Rev. B \textbf{101}, 115142 (2020).

\bibitem{pav23} G. A. P. Ribeiro and G. Rigolin, Phys. Rev. A \textbf{107}, 052420 (2023); 
G. A. P. Ribeiro and G. Rigolin, Phys. Rev. A \textbf{109}, 
012612 (2024).

\bibitem{pav24}  G. A. P. Ribeiro and G. Rigolin, 
e-print arXiv:2403.10193  [quant-ph].

\bibitem{oll01}  H. Ollivier and W. H. Zurek, Phys. Rev. Lett. \textbf{88}, 017901 (2001).

\bibitem{hen01}  L. Henderson and V. Vedral, J. Phys. A: Math. Gen. \textbf{34}, 6899 (2001).

\bibitem{wig63} E. P. Wigner and M. M. Yanase, 
Proc. Natl. Acad. Sci. U.S.A. \textbf{49}, 910 (1963).

\bibitem{ben93} C. H. Bennett, G. Brassard, C. Crepeau, R. Jozsa, A. Peres, and
W. K. Wootters, Phys. Rev. Lett. 70, \textbf{1895} (1993).

\bibitem{yeo02} Y. Yeo, Phys. Rev. A \textbf{66}, 062312 (2002).

\bibitem{rig17} R. Fortes and G. Rigolin, Phys. Rev. A \textbf{96}, 022315 (2017).

\bibitem{mal16} A. L. Malvezzi, G. Karpat, B. \c{C}akmak, F. F. Fanchini,
T. Debarba, and R. O. Vianna, Phys. Rev. B \textbf{93}, 184428 (2016).

\bibitem{hua14} Y. Huang, New J. Phys. \textbf{16}, 033027 (2014).

\bibitem{cil10} L. Ciliberti, R. Rossignoli, and N. Canosa, 
Phys. Rev. A \textbf{82}, 042316 (2010).

\bibitem{hua13}  Y. Huang, Phys. Rev. A \textbf{88}, 014302 (2013).

\bibitem{yur15} M. A. Yurischev, Quantum Inf. Process. \textbf{14}
3399 (2015). 

\bibitem{luo05} S. Luo, Phys. Rev. A \textbf{72}, 042110 (2005).

\bibitem{rig15} R. Fortes and G. Rigolin, Phys. Rev. A \textbf{92}, 012338 (2015); \textbf{93}, 062330 (2016).

\bibitem{uhl76} A. Uhlmann, Rep. Math. Phys. \textbf{9}, 273 (1976).

\bibitem{nie00} M. A. Nielsen and I. L. Chuang, \textit{Quantum Computation
and Quantum Information} (Cambridge University Press,
Cambridge, 2000).

\bibitem{gor06} G. Gordon and G. Rigolin, Phys. Rev. A \textbf{73}, 042309 (2006);
\textbf{73}, 062316 (2006); Eur. Phys. J. D \textbf{45}, 347 (2007). 

\bibitem{tos23} F. Toscano, D. G. Bussandri, G. M. Bosyk, A. P. Majtey, and M. Portesi, Phys. Rev. A \textbf{108}, 042428 (2023).

\bibitem{tos24}  D. G. Bussandri, G. M. Bosyk, and F. Toscano, 
Phys. Rev. A \textbf{109}, 032618 (2024).

\bibitem{yan66} C. N. Yang and C. P. Yang, Phys. Rev. \textbf{147}, 303 (1966).

\bibitem{clo66} J. Cloizeaux and M. Gaudin, J. Math. Phys. \textbf{7}, 1384 (1966).

\bibitem{klu92} A. Kl\"umper, Ann. Phys. \textbf{1}, 540 (1992); 
Z. Phys. B \textbf{91}, 507 (1993).

\bibitem{bor05} M. Bortz and F. G\"ohmann, Eur. Phys. J. B \textbf{46}, 399 (2005).

\bibitem{boo08} H. E. Boos, J. Damerau, F. G\"ohmann, A. Kl\"umper, J. Suzuki, and
A. Wei\ss e, J. Stat. Mech. (2008) P08010. 

\bibitem{tri10} C. Trippe, F. G\"ohmann, and A. Kl\"umper, Eur. Phys. J. B \textbf{73}, 253 (2010).

\bibitem{tak99} M. Takahashi, 
\textit{Thermodynamics of one-dimensional solvable models}
(Cambridge University Press, Cambridge, 1999).

\bibitem{lie61} E. Lieb, T. Schultz, and D. Mattis, 
Ann. Phys. \textbf{16}, 407 (1961).

\bibitem{bar70} E. Barouch, B. M. McCoy, and M. Dresden, 
Phys. Rev. A \textbf{2}, 1075 (1970).

\bibitem{bar71} E. Barouch and B. M. McCoy, 
Phys. Rev. A \textbf{3}, 786 (1971).

\bibitem{pfe70} P. Pfeuty, Ann. Phys. (New York) \textbf{57}, 79 (1970).

\bibitem{zho10} M. Zhong and P. Tong, 
J. Phys. A: Math. Theor. \textbf{43}, 505302 (2010).

\bibitem{ron15}  X. Rong, J. Geng, F. Shi, Y. Liu, K. Xu, W. Ma, F. Kong, Z. Jiang, Y. Wu, and J. Du, Nat. Commun. \textbf{6}, 8748 (2015).

\bibitem{bra19} C. E. Bradley, J. Randall, M. H. Abobeih, R. C. Berrevoets, 
M. J. Degen, M. A. Bakker, M. Markham, D. J. Twitchen, and T. H. Taminiau, 
Phys. Rev. X \textbf{9}, 031045 (2019).

\bibitem{xie19} T. Xie, Z. Zhao, X. Kong, W. Ma, M. Wang, X. Ye, P. Yu, Z. Yang, S. Xu, P. Wang \textit{et al.}, 
Sci. Adv. \textbf{7}, eabg9204 (2021).

\bibitem{noi22} A. Noiri, K. Takeda, T. Nakajima, T. Kobayashi, 
A. Sammak, G. Scappucci, and S. Tarucha, 
Nature \textbf{601}, 338 (2022).

\bibitem{xue22} X. Xue, M. Russ, N. Samkharadze, B. Undseth, 
A. Sammak, G. Scappucci, and L. M. Vandersypen, 
Nature \textbf{601}, 343 (2022).

\bibitem{mad22} M. T. M{\k{a}}dzik, S. Asaad, A. Youssry, 
B. Joecker, K. M. Rudinger, E. Nielsen, K. C. Young, 
T. J. Proctor, A. D. Baczewski, A. Laucht \textit{et al.}, 
Nature \textbf{601}, 348 (2022).

\bibitem{xie22} T. Xie, Z. Zhao, S. Xu, X. Kong, Z. Yang, M. Wang,
Y. Wang, F. Shi, and J. Du, 
Phys. Rev. Lett. \textbf{130}, 030601 (2023).

\bibitem{gam17} J. M. Gambetta, J. M. Chow, and M. Steffen, 
npj Quantum Inf. \textbf{3}, 2 (2017).

\bibitem{ohn99} Y. Ohno, R. Terauchi, T. Adachi, F. Matsukura, and 
H. Ohno, Phys. Rev. Lett. \textbf{83}, 4196 (1999).

\bibitem{han03} R. Hanson, B. Witkamp, L. M. K. Vandersypen, 
L. H. W. van Beveren, J. M. Elzerman, and L. P. Kouwenhoven, 
Phys. Rev. Lett. \textbf{91}, 196802 (2003).

\bibitem{zin10} A. F. Zinovieva, A. V. Dvurechenskii, N. P. Stepina, 
A. I. Nikiforov, and A. S. Lyubin,
Phys. Rev. B \textbf{81}, 113303 (2010).

\bibitem{pet22} L. Petit, M. Russ, G. H. G. J. Eenink, W. I. L. Lawrie,
J. S. Clarke, L. M. K. Vandersypen, and M. Veldhorst, 
Commun. Mater. \textbf{3}, 82 (2022).


\end{thebibliography}
\end{document}